\documentclass{jpp}
\usepackage{graphicx}
\usepackage{epstopdf, epsfig}
\usepackage{hyperref}
\usepackage{color}
\usepackage{bm}
\usepackage{csquotes}
\usepackage{subcaption}
\newcommand{\refeq}[1]{(\ref{#1})}
\newcommand{\beqn}{\begin{eqnarray}}
\newcommand{\eeqn}{\end{eqnarray}}
\newcommand{\xp}{\times}
\newcommand{\fract}[2]{#1 / #2 }
\newcommand{\mb}[1]{\bm{#1}}
\newcommand{\imag}{{\rm{i}}}
\newcommand{\ltsp}{c}
\newcommand{\bu}{\mb{b}}
\newcommand{\bvec}{\mb{B}}
\newcommand{\evec}{\mb{E}}
\newcommand{\bmag}{B}
\newcommand{\spe}{p}
\newcommand{\fldl}{\alpha}
\newcommand{\flxl}{\psi}
\newcommand{\kpara}{k_\|}
\newcommand{\sigkp}{\sigma_{k_\|}}
\newcommand{\rr}{{\mb{r}}}
\newcommand{\gc}{{\mb{R}}}
\newcommand{\ttime}{t}
\newcommand{\ts}{{t_{\rm{s}}}}
\newcommand{\tf}{{t}}
\newcommand{\nbl}{\nabla}
\newcommand{\nblf}{\nbl}
\newcommand{\nbls}{\nbl}
\newcommand{\pvel}{\mb{v}}
\newcommand{\gyrophase}{\vartheta}
\newcommand{\energy}{\varepsilon}
\newcommand{\pitch}{\lambda}
\newcommand{\sign}{\sigma}
\newcommand{\vpar}{v_\|}
\newcommand{\vmag}{v}
\newcommand{\lpar}{\theta}
\newcommand{\arcl}{z}
\newcommand{\kpar}{\bu \cdot \nbl \lpar}
\newcommand{\lscal}{a}
\newcommand{\saffac}{q}
\newcommand{\esav}[1]{\langle{#1}\rangle^{{\scriptsize\rm{ES}}}}
\newcommand{\gav}[2]{\langle{#1}\rangle^{\gyrophase}_{}}
\newcommand{\flav}[1]{\left\langle{#1}\right\rangle_{\lpar}}
\newcommand{\intv}[1]{\int \! d^3 \mb{v} #1 }
\newcommand{\ovl}[1]{\overline{#1}}
\newcommand{\tld}[1]{\tilde{#1}}
\newcommand{\ptl}[1]{\phi_{#1}} 
\newcommand{\iptl}[1]{\ovl{\phi}_{#1}} 
\newcommand{\eptl}[1]{\tld{\phi}_{#1}} 
\newcommand{\egptle}[1]{\tld{\varphi}_{ #1}} 
\newcommand{\ihhe}[1]{\ovl{h}_{#1}} 
\newcommand{\ehhe}[1]{\tld{h}_{#1}} 
\newcommand{\igge}[1]{\ovl{g}_{ #1}} 
\newcommand{\egge}[1]{\tld{g}_{ #1}} 
\newcommand{\dlf}{{\delta \! f}}
\newcommand{\idlf}{{\ovl{\delta \! f}}}
\newcommand{\edlf}{{\widetilde{\delta \! f}}}
\newcommand{\idlfe}{{\ovl{\delta \! f}_{\rm{e}}}}
\newcommand{\edlfe}{{\widetilde{\delta \! f}_{\rm{e}}}}
\newcommand{\eqlbe}{F_{0}}
\newcommand{\gyrds}{\rho_{{\rm{th}},\spe}}
\newcommand{\gyrdi}{\rho_{\rm{th,i}}}
\newcommand{\gyrde}{\rho_{\rm{th,e}}}
\newcommand{\dens}{n}
\newcommand{\dense}{n_{0\rm{e}}}
\newcommand{\edensi}{\tld{n}_{\rm{i}}}
\newcommand{\edense}{\tld{n}_{\rm{e}}}
\newcommand{\edenss}{\tld{n}_{\spe}}
\newcommand{\temp}{T}
\newcommand{\temps}{T_\spe}
\newcommand{\tempi}{T_{\rm{i}}}
\newcommand{\tempe}{T_{\rm{e}}}
\newcommand{\vtheri}{v_{\rm{th,i}}}
\newcommand{\vthere}{v_{\rm{th,e}}}
\newcommand{\vthers}{v_{{\rm{th}},\spe}}
\newcommand{\zeds}{Z_\spe}
\newcommand{\zedi}{Z_{\rm{i}}}
\newcommand{\charge}{e}
\newcommand{\drv}[2]{\frac{\partial #1}{\partial #2}}
\newcommand{\drvt}[2]{{\partial #1}/{\partial #2}}
\newcommand{\mass}{m}
\newcommand{\ma}{m_\spe}
\newcommand{\me}{m_{{\rm{e}}}}
\newcommand{\mi}{m_{{\rm{i}}}}
\newcommand{\massrt}{\left(m_{\rm{e}}/m_{\rm{i}}\right)^{1/2}}
\newcommand{\massruct}{\left({m_{\rm{i}}}/{m_{\rm{e}}}\right)^{3/2}}
\newcommand{\vme}{\mb{v}_{M}}
\newcommand{\evee}{\tld{\mb{v}}_{E}}
\newcommand{\ivee}{\ovl{\mb{v}}_{E}}
\newcommand{\cycfs}{\Omega_\spe}
\newcommand{\cycfe}{\Omega_{\rm{e}}}
\newcommand{\ecope}{\widetilde{C_{\rm{}}}}
\newcommand{\cfreqii}{\nu_{\rm{ii}}}
\newcommand{\cfreqee}{\nu_{\rm{ee}}}
\newcommand{\cfreqss}{\nu_{\spe\spe}}
\newcommand{\exbtext}{ {\evec \xp \bvec}}
\newcommand{\ifield}{\ovl{\chi}}
\newcommand{\efield}{\tld{\chi}}
\newcommand{\field}{\chi}
\newcommand{\shat}{\hat{s}}
\newcommand{\radial}{x}
\newcommand{\radials}{x_{\rm{s}}}
\newcommand{\binormal}{y}
\newcommand{\binormals}{y_{\rm{s}}}
\newcommand{\kky}{k_y}
\newcommand{\kkx}{k_x}
\newcommand{\ekky}{k_y}
\newcommand{\ekkx}{k_x}
\newcommand{\ekkvec}{\mb{k}}
\newcommand{\noy}{n_y}
\newcommand{\nox}{n_x}
\newcommand {\thetazh} {\hat{\theta}_0}
\newcommand{\ikkymax}{k_y^{\scriptsize\rm{max}}}
\newcommand{\ikkxmax}{k_x^{\scriptsize\rm{max}}}
\newcommand{\negrid}{n_\varepsilon}
\newcommand{\npitch}{n_\lambda}
\newcommand{\npitchp}{n_{\lambda,\rm{p}}}
\newcommand{\npitcht}{n_{\lambda,\rm{t}}}
\newcommand{\nlpar}{n_\lpar}
\newcommand{\saffacz}{\saffac_0}
\newcommand{\bscal}{B_{\scriptsize \rm{ref}}}
\newcommand{\rhominor}{\rho}
\newcommand{\rhominorz}{\rho_{0}}
\newcommand{\rminorz}{r_0}
\newcommand{\dpsidx}{d \flxl / d \radial}
\newcommand{\daldy}{d \fldl / d \binormal}
\newcommand{\kkxnorm}{\bscal \lscal \rhominorz/ \saffacz}
\newcommand{\kkynorm}{\bscal \lscal \drvt{\rhominor}{\flxl}|_{\rhominorz}}
\newcommand{\kxfac}{\kappa_\rho}
\newcommand{\kxfacdef}{\bscal(\fract{d \radial}{d\flxl} )(\fract{d \binormal}{d \fldl})}
\newcommand{\whate}{\hat{\omega}_{E}}
\newcommand{\growth}{\gamma}
\newcommand{\growthmaxcbc}{\gamma^{\scriptsize \rm{CBC}}_{\scriptsize \rm{max}}}
\newcommand{\growthmax}{\gamma_{\scriptsize \rm{max}}}

\newcommand{\ispos}{(\radials, \binormals)}
\newcommand{\ispot}{(\ts,\radials,\binormals)}
\newcommand{\avgamma}{\langle\growth\rangle_{\ts,\radials,\binormals}}
\newcommand{\iwe}{\ovl{\omega}_{E \times B}}
\newcommand{\wm}{\omega_M}
\newcommand{\lti}{L_{\tempi}}
\newcommand{\ltt}{L_{\temp}}
\newcommand{\lte}{L_{\tempe}}
\newcommand{\lteeff}{L_{\tempe}^{\rm{eff}}}
\newcommand{\rstari}{\rho^{\ast}_{ \rm{th,i} }}
\newcommand{\rmaj}{R}
\newcommand{\rmajref}{R_{\scriptsize\rm{ref}}}
\newcommand{\rmajmin}{R_{\scriptsize\rm{min}}}
\newcommand{\rmajmax}{R_{\scriptsize\rm{max}}}
\newcommand{\wstarn}{\omega^{\ast}_n}
\newcommand{\wstart}{\omega^{\ast}_{T_{\rm{e}}}}
\newcommand{\wstarphi}{\ovl{\omega}^{\ast}_{\phi}}
\newcommand{\wstardn}{\ovl{\omega}^{\ast}_{\delta\! n,h}}
\newcommand{\wstardt}{\ovl{\omega}^{\ast}_{\delta\! T_{\rm{e}}}}
\newcommand{\lln}{L_n}
\newcommand{\llneff}{L_{n}^{\rm{eff}}}
\newcommand{\llneffh}{L_{n,h}^{\rm{eff}}}
\newcommand{\igammaest}{\gamma_{\scriptsize\rm{ITG}}}
\newcommand{\egammaest}{\gamma_{\scriptsize\rm{ETG}}}
\newcommand{\ikkest}{k_{\scriptsize\rm{ITG}}}
\newcommand{\ekkest}{k_{\scriptsize\rm{ETG}}}
\newcommand{\ing}{\ovl{\delta \! n}}
\newcommand{\inh}{\ovl{\delta \! n}_h}
\newcommand{\ith}{\ovl{\delta \! \tempe}}

\begin{document}

\bibliographystyle{jpp}
 
\title{  Stabilisation of short-wavelength instabilities
 by parallel-to-the-field shear in long-wavelength $\exbtext$ flows}
\author{M. R. Hardman\aff{1,2}
  \corresp{\email{michael.hardman@physics.ox.ac.uk}},
  M. Barnes\aff{1,2}, \and C. M. Roach\aff{2} }

\affiliation{
\aff{1}Rudolf Peierls Centre for Theoretical Physics, University of Oxford, Oxford OX1 3PU, UK
\aff{2}Culham Centre for Fusion Energy, UKAEA, Abingdon OX14 3DB, UK
}
\date{\today}

\maketitle

\begin{abstract}
 Magnetised plasma turbulence can have a multiscale character: instabilities driven by
 mean temperature gradients drive turbulence at the disparate scales of the ion and
 the electron gyroradii.
 Simulations of multiscale turbulence, using
 equations valid in the limit of
 infinite scale separation, reveal novel cross-scale interaction mechanisms in these plasmas.
 In the case that
 both long-wavelength (ion-gyroradius-scale) and short-wavelength (electron-gyroradius-scale)
 linear instabilities 
 are driven far from marginal stability,
 we show that the short-wavelength instabilities are
 suppressed by interactions with long-wavelength turbulence.
 The observed suppression is a result of two effects:
 parallel-to-the-field-line shearing by the long wavelength $\exbtext$ flows,
 and the modification of the background density gradient by long-wavelength fluctuations.
 In contrast, simulations of multiscale turbulence where instabilities at both scales
 are driven near marginal stability demonstrate that 
 when the long-wavelength turbulence is sufficiently collisional and zonally dominated
 the effect of cross-scale interaction can be parameterised solely in terms of
 the local modifications to the mean density and temperature gradients. 
 We discuss physical arguments that qualitatively explain how a change
 in equilibrium drive leads to the observed transition in the impact of the cross-scale interactions. 

\end{abstract}

\section{Introduction}\label{sec:intro}
 In a magnetised plasma, gradients in the mean temperature
of the component particle species act as sources
 of free energy that drive instability.
 Instabilities saturate through nonlinear interactions to form turbulence;
the nature of the turbulence is determined by the character of the underlying instabilities. 
In this paper we consider the effect of cross-scale nonlinear interactions
 in turbulence driven at the well-separated space-time scales
 associated with ion and electron dynamics, respectively. 
 This represents a fundamentally different 
 system to the one usually employed in turbulence studies,
 which consists of a single injection range, a single inertial range, and a single dissipation range.

 Instabilities in magnetised plasmas typically have a structure that is elongated along field
lines; this is a result of the Lorentz force which causes particles to perform gyro orbits in the
 plane perpendicular to the magnetic field line,
 whilst allowing unimpeded motion in the parallel-to-the-field direction.
By definition, in a magnetised plasma the typical
 particle gyroradius is much smaller than the device scale $\lscal$ that 
 determines the parallel-to-the-field scale of instabilities.   
 In the core of magnetic confinement fusion devices two of the dominant instabilities which
 drive turbulence are the ion temperature gradient (ITG) and the electron temperature gradient 
(ETG) instabilities \cite{cowleyPoFB91,RomanelliITG,HortonETG,LeeETG}. 
 The ITG instability drives turbulence at the scale of the
 ion thermal gyroradius $\gyrdi$, at frequencies of order the ion transit frequency
 $\vtheri / \lscal$, with $\vtheri$ the ion thermal speed.
 The ETG instability 
 drives turbulence at the scale of the
 electron thermal gyroradius $\gyrde$, at frequencies of order the electron transit frequency
 $\vthere / \lscal$, with $\vthere$ the electron thermal speed.
 
 When the ion and electron temperatures are approximately equal
 the space and time scales associated with the ETG and ITG instabilities can be well separated: 
 the separation of  scales is determined by the square root of the electron-to-ion mass ratio 
 $\massrt \sim \gyrde/\gyrdi \sim \vtheri/\vthere$. 
 For the deuterium ions commonly used in magnetic confinement fusion,
 the small value of $\massrt \approx 1/ 60$ 
 allows for the possibility that two distinct types of turbulence co-exist 
 at disparate space-time scales. 
 This possibility has sparked considerable interest in multiscale turbulence,
 see e.g. 
 \cite{Maeyama_2017_NF,maeyama2017supression,maeyama2015cross,howard2016enhanced,howard2016comparison,howard2015fidelity,howard2014synergistic,Bonanomi_2018_ImpactofES,gorler2008scale,candy2007effect,waltz2007coupled,2016StaeblerMultiscale,2017StaeblerMultiscale,Creely_2019,Itoh_2001}. 
 
 Direct numerical simulations (DNS) of multiscale plasma turbulence show that 
 the interactions between long-wavelength and short-wavelength
 scales can be important for determining the character of the turbulence 
 and the flux of heat exhausted
 from the fusion plasma \cite{howard2016enhanced,howard2016comparison,maeyama2015cross,Maeyama_2017_NF}. 
 The small value of $\massrt$ makes 
 DNS of multiscale plasma turbulence extremely challenging;
 the increased cost of multiscale DNS compared to conventional simulations 
 scales with $\massruct$. Recent multiscale DNS have been performed using
 the electron-to-deuterium mass ratio (e.g. \cite {howard2016enhanced,howard2016comparison, Bonanomi_2018_ImpactofES}) and electron-to-hydrogen
 mass ratio (e.g. \cite {maeyama2015cross,Maeyama_2017_NF}).
 The multiscale DNS provide evidence showing
 that the presence of short-wavelength turbulence
 can sometimes modestly enhance long-wavelength, ITG-driven fluxes
 \cite{howard2016enhanced,howard2016comparison,maeyama2015cross,Maeyama_2017_NF}.
 In contrast, multiscale DNS of microtearing mode (MTM) driven turbulence show
 that the presence of short-wavelength ETG modes can suppress
 the long-wavelength MTM \cite{maeyama2017supression}.
 Most relevant to the results presented in this paper,
 the multiscale DNS give clear evidence that shows that short-wavelength turbulence can be 
 suppressed in the presence
 of long-wavelength turbulence as a result of cross-scale interaction 
 \cite {maeyama2015cross,Maeyama_2017_NF,howard2016comparison}.                                                          
 
In this paper we use numerical simulations of scale-separated turbulence 
to demonstrate the effect of long-wavelength turbulence on short-wavelength fluctuations.
 We show that strongly driven, long-wavelength 
 turbulence 
 can stabilise the short-wavelength ETG instability through cross-scale interaction.
 We show that this stabilisation is due to two effects: the modification of the background
 drives of instability by gradients of long-wavelength fluctuations; and
 the parallel-to-the-field 
shearing of short-wavelength fluctuations by long-wavelength $\exbtext$ drifts. 
These mechanisms may explain the suppression of the short-wavelength modes observed in DNS
of strongly driven turbulence. 
We show that parallel-to-the-field shearing can be
 a significant cross-scale interaction mechanism far above marginal stability,
 but that it can be less important in
turbulence driven near marginal stability. We examine one
 example of near-marginal turbulence where we find that the effect of
 cross-scale interaction on the ETG instability can be described with the local
 modification of the density and temperature gradient length scales by gradients of long-wavelength fluctuations. 

 \section{Scale-separated model }\label{sec:Scale-separated}
 The local, $\dlf$ gyrokinetic equations describe
 the evolution of magnetised plasma turbulence
 driven by mean gradients 
 in the limit that $\gyrdi/\lscal \rightarrow 0$ and $\gyrde/\lscal \rightarrow 0$
 \cite{cattoPP78,friemanPoF82,brizardRMP07}; 
 this model is the starting point in the derivation of
 scale-separated equations for multiscale, electrostatic turbulence \cite{hardmanpaper1}. 
 We now briefly review this derivation, first presented in \cite{hardmanpaper1}.
 Ultimately, we find that the local gradients of 
 slowly evolving, long-wavelength turbulence 
 act to modify the mean gradients
 and flows which drive (or suppress) rapidly evolving, short-wavelength  fluctuations. 
 In analogy to the coupling between turbulence and large-scale transport \cite{sugamaPoP97,abelRPP13},
 we might expect short-wavelength turbulence to generate fluxes that diffuse long-wavelength
 turbulence. However, this effect is small by $\massrt$ in the ordering, and hence
 too small to appear at leading order in the $\massrt$ expansion that we employ. 
 
 In the limit that $\massrt \rightarrow 0$, multiscale turbulence can be meaningfully decomposed into 
 components: the long-wavelength \enquote{ion scale} (IS) and the
 short-wavelength \enquote{electron scale} (ES). 
 We assume that IS turbulence varies only on $\gyrdi$
 perpendicular-to-the-field scales in space and on $\lscal/\vtheri$ scales in time,
 whereas ES turbulence has much finer $\gyrde$ perpendicular-to-the-field structures,
 and evolves on the rapid $\lscal / \vthere$ time scale.
 Both IS and ES turbulence have parallel-to-the-field scales of order $\lscal$.
 Here, $\gyrds=\vthers/\cycfs$, with $\spe$ the particle species index,
 $\vthers = \sqrt{2\temps/\ma}$, 
 $\cycfs = \zeds \charge \bmag /  \ma \ltsp$ the species cyclotron frequency,
 $\temps$ the species temperature, 
$\zeds$ the species charge number, $\charge $ the unit charge, $\bmag$ the magnetic field
 strength, and $\ltsp$ the speed of light.
 The distribution function $\dlf$
 and electrostatic potential $\ptl{}$  of the turbulence are sums of the IS and ES components; 
 \beqn \dlf = \idlf + \edlf, \quad \ptl{} = \iptl{} + \eptl{}, \label{eq:dfnsplit}\eeqn
 where $\ifield$ and $\efield$ are the IS and ES pieces of any quantity $\field$, respectively.
 To carry out this decomposition, we introduce the ES average $\esav{\cdot}$, an average over $\gyrde$
 perpendicular-to-the-field scales and $\lscal / \vthere$ times. We assume that ES turbulence
 is statistically periodic on $\gyrde$ 
 scales, i.e., $\esav{\edlf} = 0$.
 This allows us to use $\esav{\cdot}$ to extract the scale-separated, coupled gyrokinetic equations
 for $\idlf$ and $\edlf$ from the gyrokinetic equations for $\dlf$. 
 
 The IS component of the turbulence is evolved with the usual gyrokinetic
 equation for the ion species, 
 and a parallel-orbit-averaged equation
 for the electron species. 
 The IS turbulence evolves independently
 of the 
 ES turbulence to leading order in the $\massrt$ expansion. 
 At the ES  the ion species have a Boltzmann response. 
 The evolution of the ES turbulence is governed by
 the short-wavelength electron gyrokinetic equation \cite{hardmanpaper1}
 \beqn   \left(\drv{}{\tf} + \ivee\cdot \nbl\right)\egge{} +  \vpar \kpar \drv{\egge{}}{\lpar}  +   (\vme +\evee) \cdot \nblf \egge{}  + \evee \cdot (\nbl \eqlbe + \nbls \igge{}) \nonumber\eeqn
  \beqn    \hspace*{4cm} =\frac{\charge  \eqlbe}{\tempe}\left(\vpar  \kpar \drv{\egptle{}}{\lpar} +\vme \cdot \nblf \egptle{}\right)  + \ecope,
  \label{equation:electronelectronrealgg}
   \eeqn
 where we express the ES  gyrokinetic equation in terms of
 the particle guiding centre $\gc = \rr + \bu \xp \pvel / \cycfe$,
 with $\rr$ the particle position, $\bu$ the magnetic field direction,
 $\pvel$ the particle velocity; 
 $\energy$ the particle kinetic energy; $\pitch = \mu / \energy$ the pitch angle,
 with $\mu$ the 
 magnetic moment;
 $\sign$ the sign of the particle velocity in the magnetic field direction  $\vpar= \pvel\!\cdot\!\bu = \sign (2\energy/\ma)^{1/2}(1-\pitch \bmag)^{1/2}$;
 the gyroaveraged ES potential $\egptle{} = \gav{\eptl{}}{{\rm{e}},\gc}$, 
 with $\gyrophase$ the angle of gyration about $\bu$ 
 and $\gav{\cdot}{{\rm{e}},\gc}$ the electron gyroaverage
 at fixed 
 $\gc$; 
 the ES gyroaveraged electron distribution function
 $\egge{} = \gav{\edlfe}{{\rm{e}},\gc} = \ehhe{} + \charge \egptle{} \eqlbe / \tempe$, with $\edlfe(\rr) = \ehhe{}(\gc) + \charge \eptl{}(\rr) \eqlbe / \tempe$, 
 $\ehhe{}$ the non-Boltzmann part of the ES electron distribution function, 
and $\eqlbe$ the Maxwellian electron mean distribution function.
 The poloidal angle
 $\lpar$ is the parallel-to-the-field coordinate; $\vme$ is the electron magnetic drift;
 and $\ecope$ accounts for the effects of collisions on $\egge{}$.
 The long-wavelength and short-wavelength  $\exbtext$ drifts are $\ivee = (\ltsp/\bmag) \bu \xp \nbls \iptl{} $
 and  $\evee = (\ltsp/\bmag) \bu \xp \nblf \egptle{} $, respectively; and
  $\igge{} = \idlfe$
 is the long-wavelength, IS, electron distribution function. Gyroaverages do not appear in the definitions of 
 $\ivee$ or $\nbls \igge{}$ as $\gyrde$ is much smaller than the spatial scale of the IS structures.  
 Equation \refeq {equation:electronelectronrealgg} is closed by the ES quasineutrality relation
 $\zedi \edensi = \edense$, with $\edenss$ the ES fluctuating density of species $\spe$.
 The ES quasineutrality relation can be written
 \beqn  \intv{|_{\rr}} \left( \egge{}(\gc) + \frac{\charge \eqlbe}{\tempe}\left(\eptl{}(\rr) - \egptle{}(\gc) \right) \right)
   = - \frac{ \zedi \charge \eptl{}(\rr)}{\tempi}\dense, \label{eq:quasineutrality}  \eeqn 
 where 
 $\dense$ is the mean electron density, 
 and the subscript $\rr$ on the volume element in
 the velocity space integral indicates that the integral
 is to be taken at fixed particle position.
 
 Cross-scale interaction is mediated by the terms
 $\ivee\! \cdot\! \nblf \egge{}$ and $\evee\! \cdot\! \nbls \igge{}$
 in the electron gyrokinetic equation, equation \refeq{equation:electronelectronrealgg}.
 The gradient 
 $\nbls\igge{}$ and the
 drift $\ivee$ are constant on $\lscal/\vthere$ time scales,
 and so appear as additional \enquote{equilibrium} terms
 in the short-wavelength electron gyrokinetic equation.
 By inspecting equation \ref{equation:electronelectronrealgg} we see that we are able to
 ascribe simple physical interpretations to the cross-scale terms:
  the term $\ivee\cdot\nbl \egge{}$
  represents an advection term that will introduce velocity shear; 
  and the gradient $\nbls\igge{}$ appears in the term $\evee\cdot(\nbl\eqlbe+\nbls\igge{})$
  i.e., $\nbls\igge{}$ modifies the usual equilibrium drive of instability $\nbls\eqlbe$.  
 Equations \refeq{equation:electronelectronrealgg}
 and \refeq{eq:quasineutrality} are solved in a thin ES flux tube embedded within a
 larger IS flux tube, see figure \ref{fig:nested_flux_tubes} for a cartoon.
 In each ES flux tube $\nbls \igge{}$ and $\ivee$ take values which
 are constant in the perpendicular-to-the-field plane; the effects of the perpendicular derivatives
 of  $\nbls \igge{}$ and $\ivee$ are small by $\massrt$ and so only appear at higher order.
 However, both $\nbls \igge{}$ and $\ivee$ can vary
 along the field line: this results in a background drive of instability
 that is no longer uniform 
 in $\lpar$, and in suppression of short-wavelength instabilities due to
 parallel-to-the-field $\exbtext$ shearing by long-wavelength flows.
 The component of $\ivee$ due to the piece of the electrostatic
 potential $\iptl{}$ that is constant on the flux surface may be removed
 from equation \refeq{equation:electronelectronrealgg} by boosting to a toroidally rotating frame \cite{mhardman2019a};
 this component of $\ivee$ can only change the instability frequency by a Doppler shift.
 We will discuss the physical interpretation of these cross-scale interaction mechanisms
 further in section \ref{sec:Interpretation}.

To enable the study of cross-scale interaction in this multiscale framework, the cross-scale terms
 $\ivee\! \cdot\! \nblf \egge{}$ and $\evee\! \cdot \! \nbls \igge{}$
 in equation \refeq{equation:electronelectronrealgg} were implemented
 in the $\dlf$ gyrokinetic code \texttt{GS2} \cite{KOTSCHENREUTHER1995CPC,mhardman2019a}.

 \begin{figure}
\begin{center}
\includegraphics[clip, trim=0cm 3.5cm 0cm 3.5cm, width=0.5\textwidth]{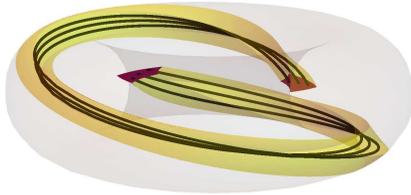}
\caption{A cartoon of a large-scale, IS flux tube
 with several narrow ES flux tubes embedded within it.} \label{fig:nested_flux_tubes}
\end{center}
\end{figure}

 \section {Numerical results. } \label{sec:Numerical}
 In this paper we  address the key question of
 how the linear stability of the ETG mode is affected by
 cross-scale interaction. We focus on a plasma in which single-scale microturbulence
 is well understood: we use a simple magnetic geometry consisting
 of concentric circular flux surfaces, with parameters
 largely corresponding to the widely-used Cyclone Base Case benchmark (CBC) \cite{DimitsCBCPoP2000}.
 In the simulations
 we take $\lscal$ to be the half-diameter of the last closed
 flux surface; 
 we take the normalised  minor radius at the centre 
 of the flux tube $\rhominorz = \rminorz/\lscal = 0.54$,
 with $\rminorz$ the minor radius of the flux surface;
 the major radius $\rmaj = 3.0 \lscal$; the safety factor at the centre 
 of the flux tube $\saffacz = 1.4$;
 the magnetic shear $\shat = \rhominor d \ln \saffac / d \rhominor = 0.8 $;
 equal ion and electron temperatures
 $\tempi = \tempe = \temp$; the normalised temperature gradient
 $\lscal/\ltt = -  d \ln \temp / d \rhominor = 2.3$; the normalised density gradient
 $ \lscal/ \lln = -  d \ln \dens / d \rhominor = 0.733$;
 the normalised self collision frequencies $\lscal \cfreqii / \vtheri = \lscal \cfreqee / \vthere = 10^{-2}$; and
 the mass ratio $\me/\mi = 1/3670$. In the IS simulations, instead of implementing
 the parallel-orbit-averaged equation for the electron species,
 we impose that the passing, nonzonal electron response is Boltzmann,
 whilst continuing to solve for the remainder of the
 electron distribution function with the usual \texttt{GS2} algorithm.
 We note that the results presented in this paper are independent of $\me/\mi$:
 IS simulations with $\me/\mi =0$ and a Boltzmann passing, nonzonal electron
 response show identical fluxes and cross-scale interactions with ETG instabilities.
 Physically, this is because the electron bounce time is shorter than the
 correlation time of the IS turbulence. 

 \begin{figure}
\begin {subfigure} {0.495\textwidth}
\begin{center}
\includegraphics[clip, trim=0cm 1.2cm 0cm 0.5cm, width=1.0\textwidth]{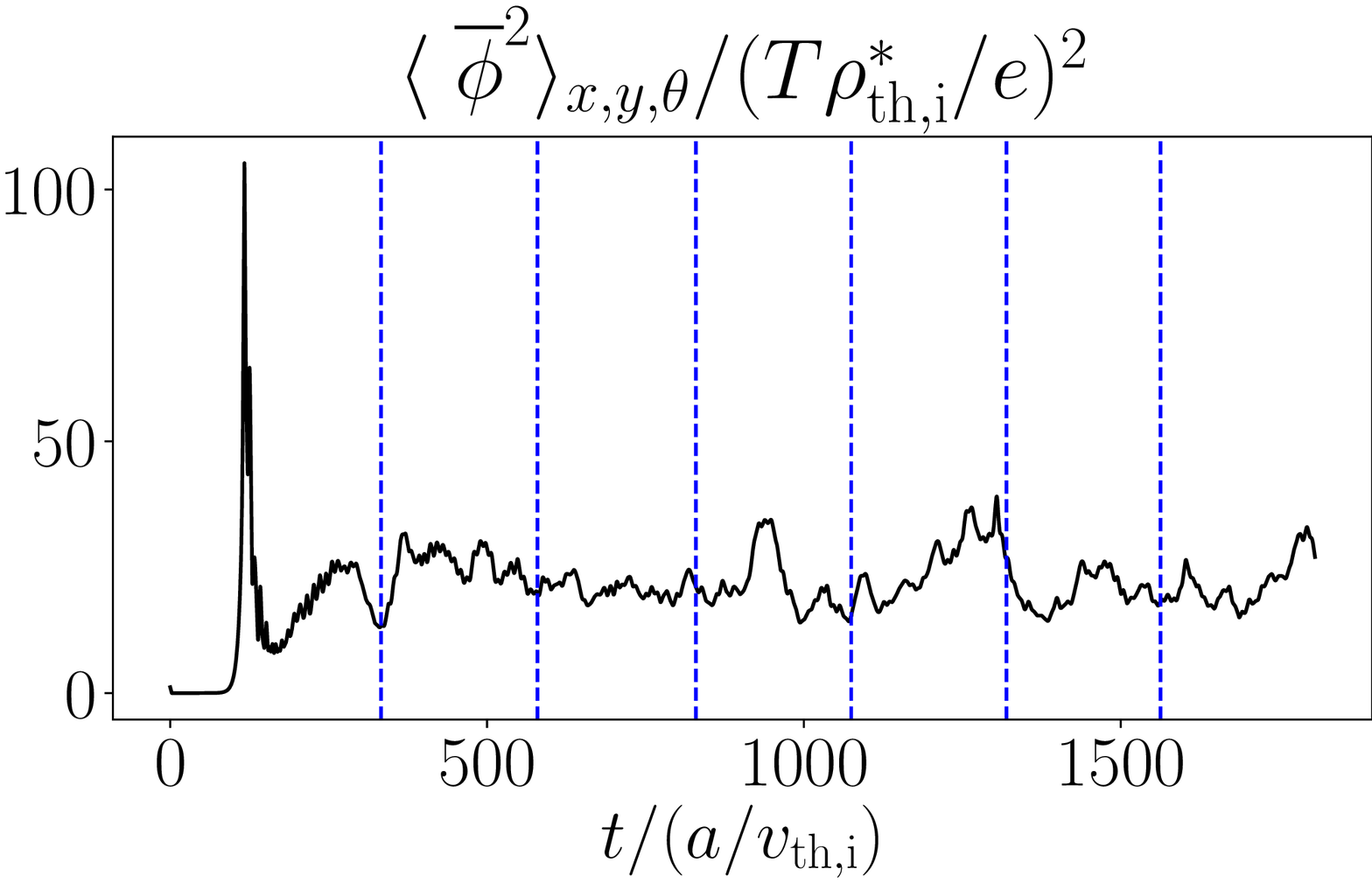}
\caption{} \label{fig:ionscale_a}
\end{center}
\end{subfigure}
\begin {subfigure} {0.495\textwidth}
\begin{center}
\includegraphics[clip, trim=0cm 1.2cm 0cm 0.5cm, width=1.0\textwidth]{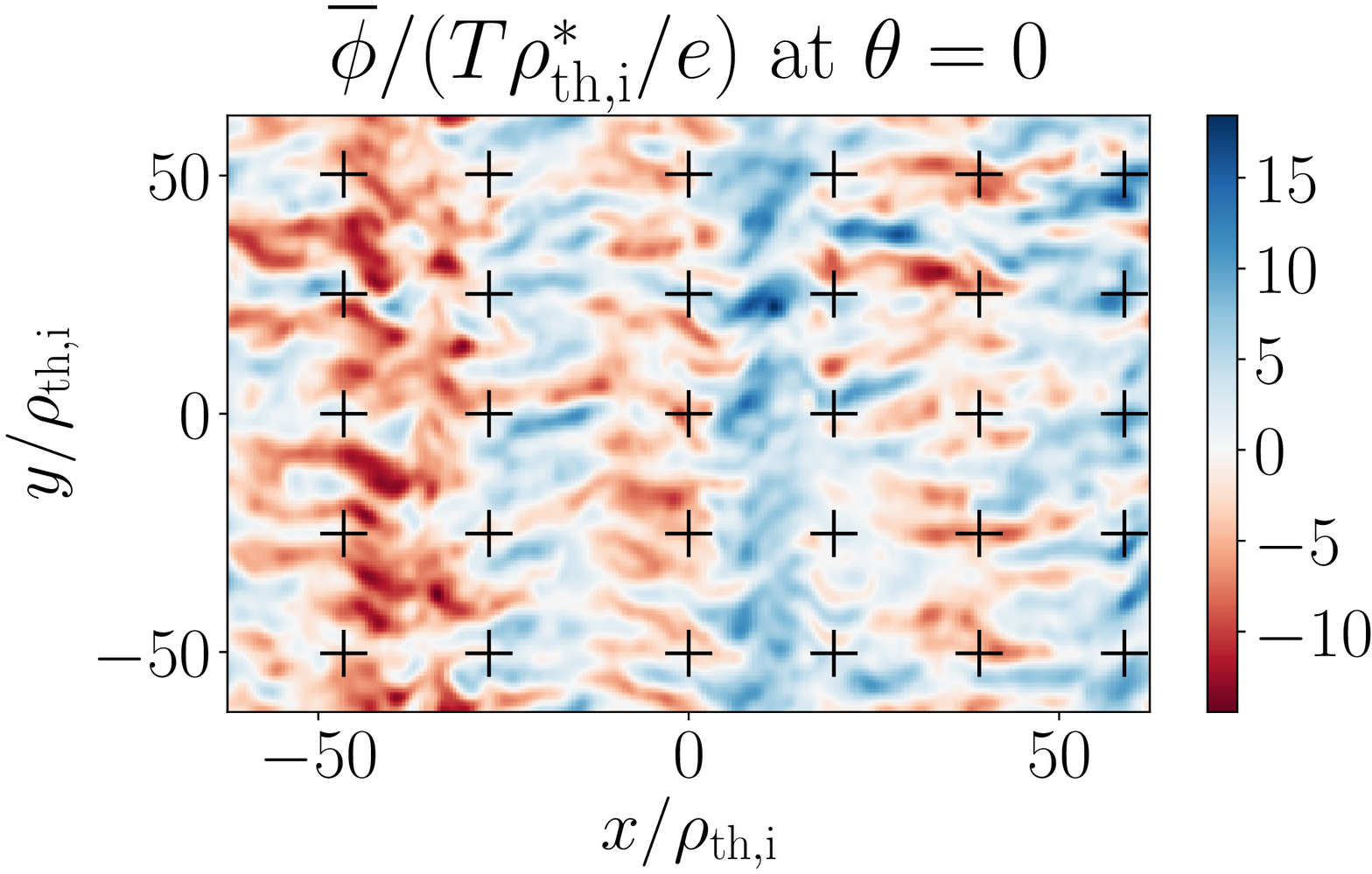}
\caption{} \label{fig:ionscale_b}
\end{center}
\end{subfigure}
\caption{ (a) The IS potential $\iptl{}$ in units of $\rstari = \gyrdi/\lscal$,
 shown as a function of time (volume-averaged),
 and (b) at the outboard midplane at $\ttime \simeq 1560 \lscal/\vtheri$.
 Times at which the IS turbulence is sampled are indicated with vertical lines,
 and sampled $\ispos$ positions are indicated by crosses.}
\label{fig:ionscale}
\end{figure}
 
 To assess the impact of cross-scale interaction on the ETG instability 
 we carry out the following numerical investigation.
 We perform simulations of
 long-wavelength ITG-driven turbulence, using
 the resolutions detailed in appendix \ref{sec:simulationresolutions:nl}.
 We compute a sample of 
 $\nbls \igge{}(\lpar,\energy,\pitch,\sign)$ and $\ivee(\lpar)$ at
 $6$ times $\ts$, on a $6\times5$ grid in radial and binormal
 position, $\radials$ and $\binormals$, respectively.
 For each $\ts$ at every sampled $\ispos$
 we compute the ETG linear growth rate $\growth\ispot$, including the cross-scale interaction terms
 $\ivee \cdot \nblf \egge{}$ and $\evee \cdot  \nbls \igge{}$ in equation
 \refeq{equation:electronelectronrealgg}.
 The resolutions used for the linear calculations are detailed
 in  appendix \ref{sec:simulationresolutions:lin}.
 This statistical approach is necessary because
 $\ivee$ and $\nbls \igge{}$ must be calculated from turbulent fields which may vary intermittently in time and space.
 Figure \ref {fig:ionscale_a} shows the spatially-averaged
 IS potential $\langle \, \iptl{}^2 \rangle_{\radial, \binormal, \lpar}(\ttime)$,
 with sampled times indicated with a vertical line; and figure \ref {fig:ionscale_b} shows
 the IS potential $\iptl{}(\radial,\binormal)$ at the outboard midplane at time
 $\ttime \simeq 1560 \lscal/\vtheri$
  -- the last sampled time in \ref {fig:ionscale_a}, with sampled positions indicated with crosses. 
   The $(\radial,\binormal)$ coordinates are proportional
to the flux-surface and field-line labels, $\flxl$ and $\fldl$, 
 respectively, that define the magnetic field $\bvec = \nbl \fldl \xp \nbl \flxl$:
  we use the perpendicular-to-the-field coordinates
  $(\radial,\binormal)$ that are defined to have units of length; i.e., 
  $\radial = (\flxl-\flxl_0) (\dpsidx)^{-1}$ and 
  $ \binormal = (\fldl-\fldl_0)  (\daldy)^{-1}$,   respectively,
  with $(\flxl_0,\fldl_0)$ the coordinates of the centre of the IS flux tube,
  $\dpsidx = \kkxnorm$ and $\daldy=\kkynorm$, where
  $\bscal$ is a reference magnetic field. We take $\bscal$ to be the toroidal magnetic field
 strength on the flux surface $\flxl_0$ at the major radius $\rmajref = (\rmajmax + \rmajmin)/2$,
 where $\rmajmax$ and $\rmajmin$ are the maximum and
 minimum major radial positions taken in the flux tube, respectively. 
  We note that we choose to define the poloidal angle $\lpar$ in
  the simulations to be such that $\kpar$ is constant in $\lpar$.    
  
\begin{figure}
\begin {subfigure} {0.495\textwidth}
\begin{center}
\includegraphics[clip, trim=0cm 1.2cm 0cm 1.05cm, width=1.0\textwidth]{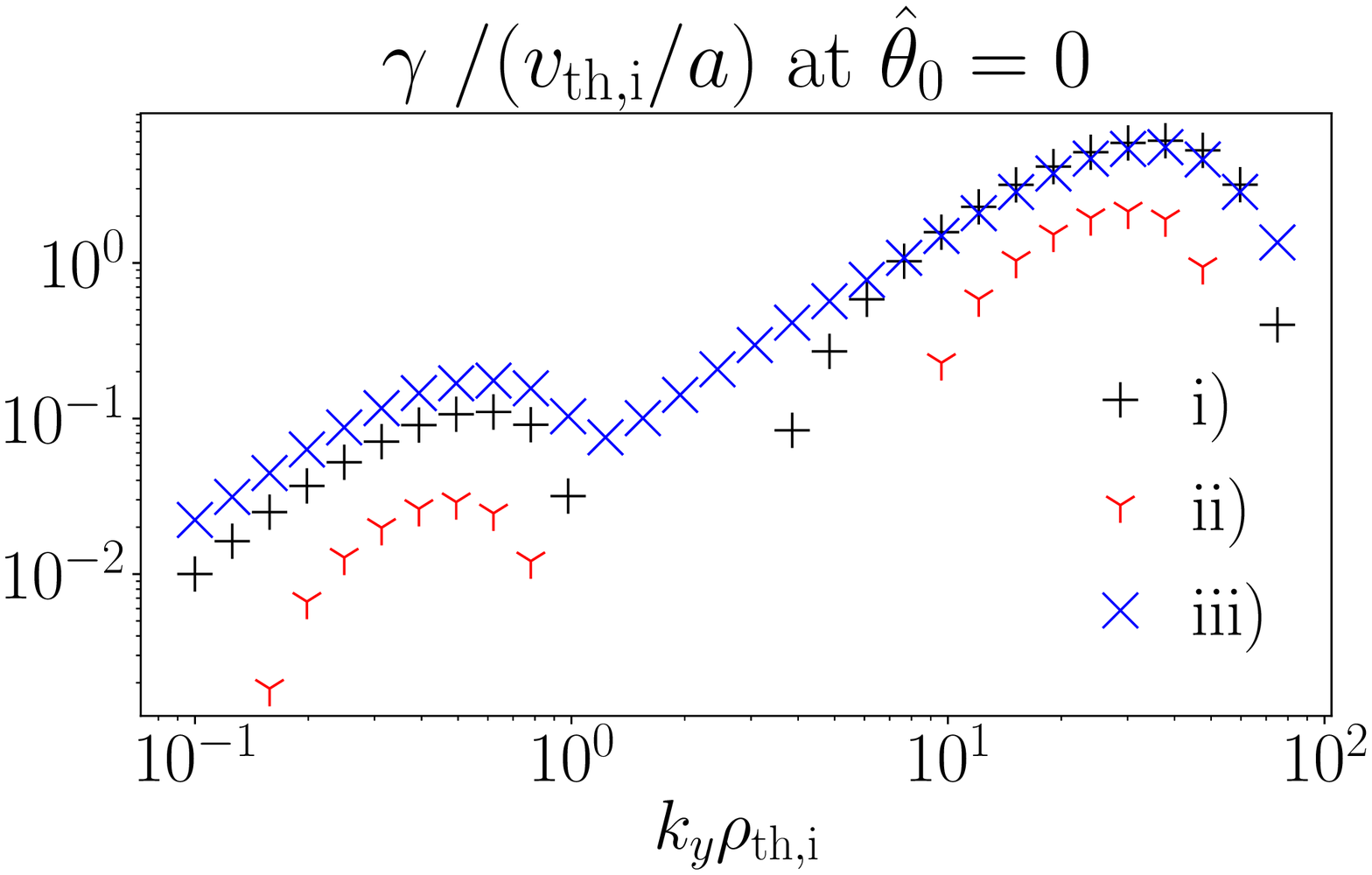}
\caption{} \label{fig:growthky}
\end{center}
\end{subfigure}
\begin {subfigure} {0.495\textwidth}
\begin{center}
\includegraphics[clip, trim=0cm 1.3cm 0cm 1.05cm, width=1.0\textwidth]{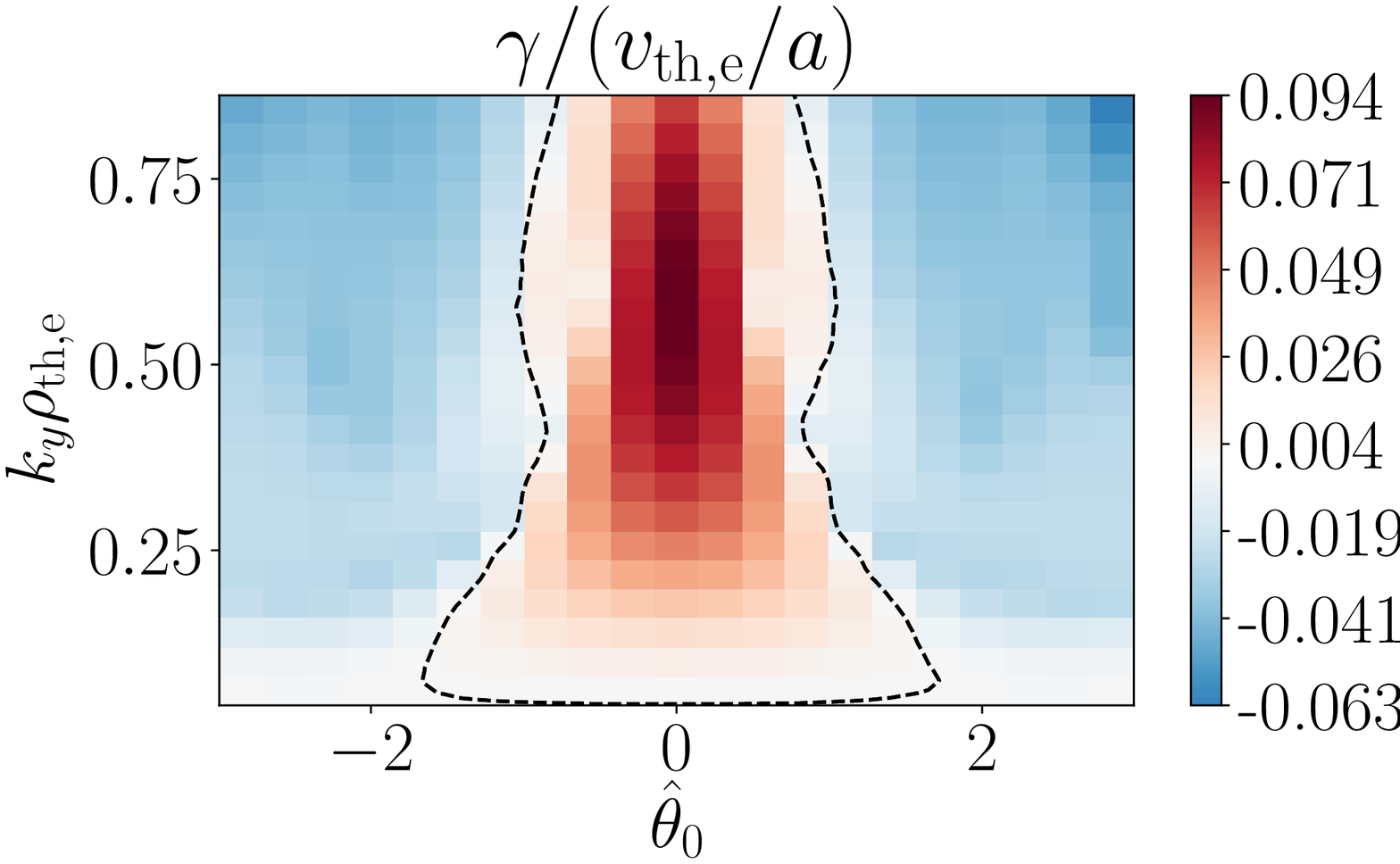}
\caption{} \label{fig:etg_noxscale}
\end{center}
\end{subfigure}
\caption{ (a) The linear growth rate for three cases:
 i) for the parameters given at the start of section \ref{sec:Numerical};
 ii) $\lscal / \lti = \lscal/\lte= 1.38$; and
 iii) $\lscal / \lti = \lscal / \lte = 2.3$ and $\lscal\cfreqss  / \vthers = 10^{-4}$.
 (b) The ETG growth rate 
 in the absence of cross-scale interaction.
 Modes within the dashed curve are unstable. } \label{fig:growth_noxscale}
\end{figure}
 Figure \ref {fig:growth_noxscale} shows the linear growth rate $\growth$ spectrum in
 the absence of cross-scale interaction. In figure \ref {fig:growthky} we
show $\growth(\ekky)$ for modes with $\thetazh =0$, where 
 $\thetazh$ is the poloidal angle at which the wave fronts of the mode align with the minor
 radial direction of the flux surface. 
 Modes with $\thetazh =0 $ have radially aligned wave fronts 
 at the outboard midplane. Figure \ref{fig:growthky} shows
 that for the simulation parameters used in this paper, see curve i), there is a natural stable gap
 in binormal wave number $\kky$ between the ITG and ETG modes:
 this serves to define the cut-off between the IS and ES.
 We note that this separation of scales
 can become larger when both the drives of instability $\lscal/\lti$ and $\lscal/\lte$ are reduced,  see curve ii),
 whereas the separation can disappear for sufficiently strong drive or low collisionality, see, e.g., curve iii).    
 In figure \ref {fig:etg_noxscale} we plot the full ETG linear growth rate spectrum 
 $\growth(\ekky,\thetazh)$ in the absence of cross-scale interaction.
 Note that the most unstable modes occur at $\thetazh =0$. 
 The dashed curve in figure \ref{fig:etg_noxscale} indicates the
 stability boundary where $\growth =0$, calculated with an interpolated spline fit.
 Figures \ref{fig:etg_xscale_a} and \ref{fig:etg_xscale_b} show the ETG growth rates 
 in the presence of cross-scale interactions 
 at two of the sampled IS positions and times. Comparing figure \ref{fig:etg_noxscale}
 with figures \ref{fig:etg_xscale_a} and \ref{fig:etg_xscale_b} shows that the effect
 of cross-scale interaction can be dramatic: in these examples
 the ETG mode is largely suppressed,
 and completely stabilised, respectively.
 
 \begin{figure}
\begin {subfigure} {0.495\textwidth}
\begin{center}
\includegraphics[clip, trim=0cm 1.3cm 0cm 1.05cm, width=1.0\textwidth]{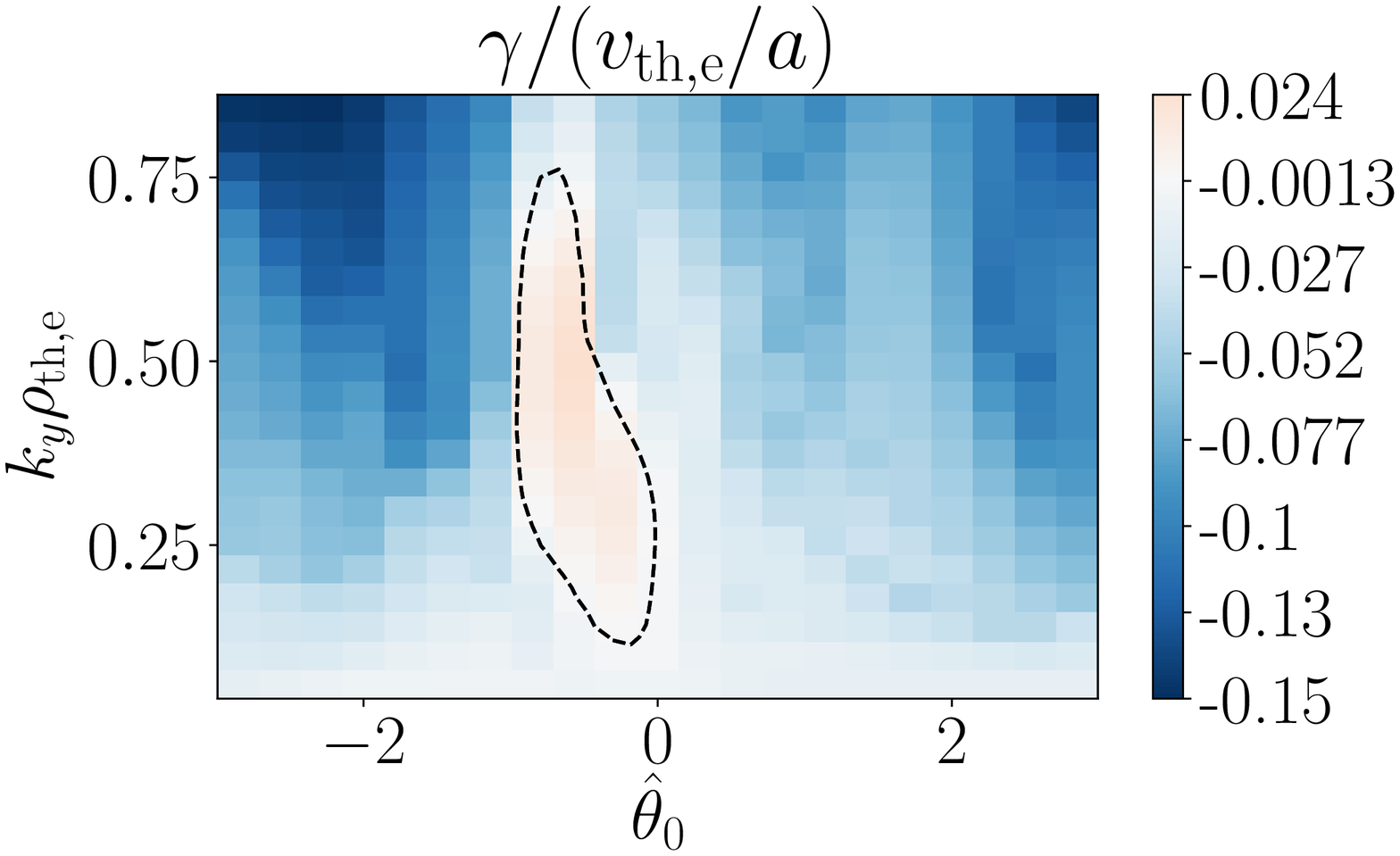}
\caption{} \label{fig:etg_xscale_a}
\end{center}
\end{subfigure}
\begin {subfigure} {0.495\textwidth}
\begin{center}
\includegraphics[clip, trim=0cm 1.3cm 0cm 1.05cm, width=1.0\textwidth]{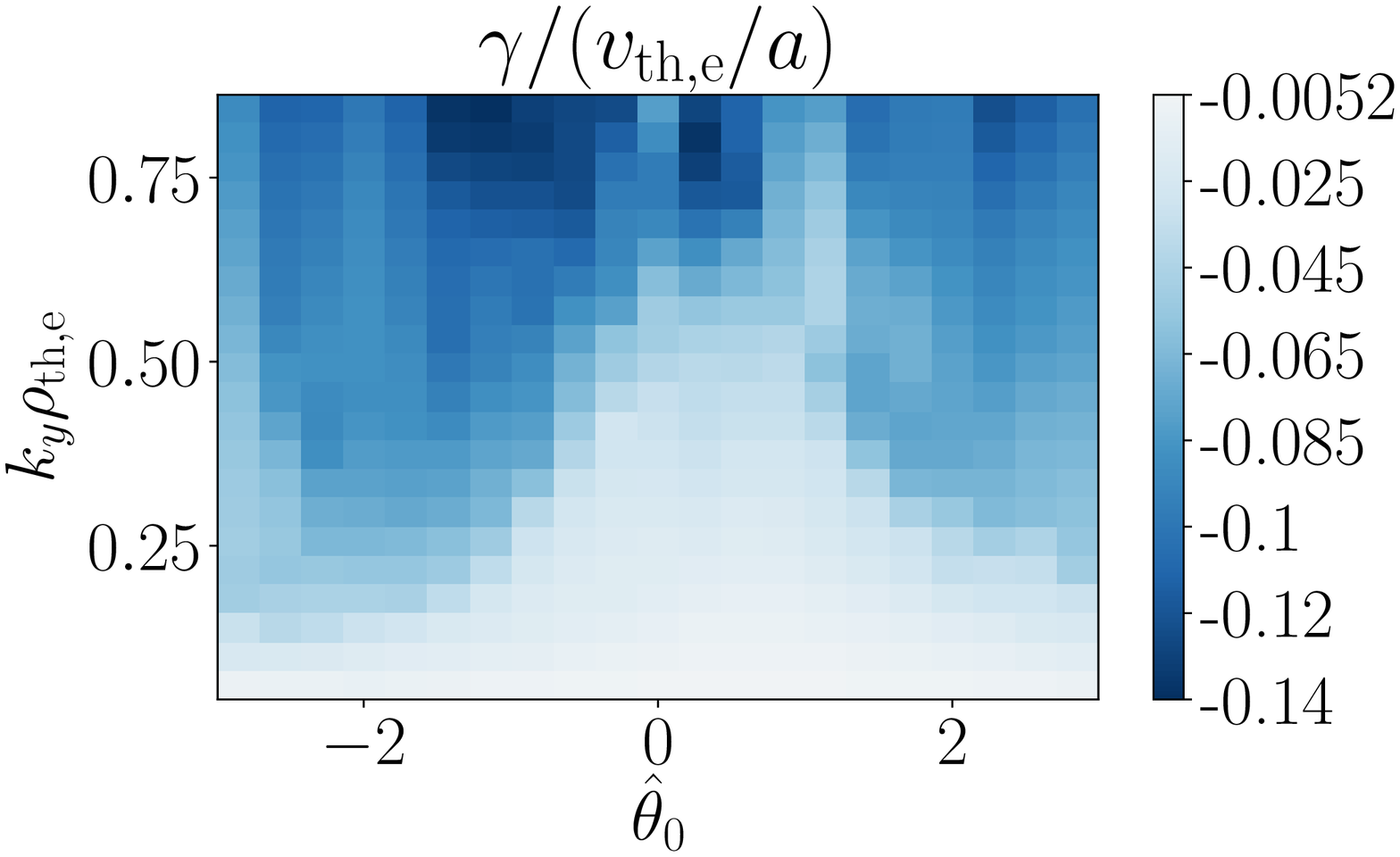}
\caption{} \label{fig:etg_xscale_b}
\end{center}
\end{subfigure}
\caption{The ETG growth rate in the presence of cross-scale interaction due to IS turbulence.
 Figures (a) and (b) show growth rates calculated at different $\ispos$ and times $\ts$.
 Note the suppression of the growth rate in (a) and (b) compared to Figure \ref{fig:etg_noxscale}. }
\label{fig:etg_theta0ky}
\end{figure}

To determine the average, or typical effect of IS turbulence on the
 ETG instability we calculate the maximum ETG growth rate $\growthmax$ at every $\ispot$,
 where we note that the maximum $\growth$ may occur at a $(\ekky,\thetazh)$ that is different from 
 the single-scale case. Figure \ref{fig:etg_growthmaxhist} shows that the majority of
 IS drifts and gradients had the effect of stabilising the ETG mode: 
 we find that  $\growthmax <0$ at $115$ $\ispot$ within the sample of $180$, with only
 a single $\ispot$ showing a $\growthmax$ larger
 than the maximum growth rate in the absence of cross-scale interaction.
 This is illustrated further in figure \ref {fig:etg_fullxscale_b},
 which shows that the average of the sampled ETG growth rates $\avgamma$ is negative, 
 for all $(\ekky,\thetazh)$. In figure \ref{fig:etg_theta0ky} we show the distribution of $\ekky$
 and $\thetazh$ of the fastest growing ETG modes 
 in the presence of IS turbulence, with only modes that have $\growthmax > 0$ included in the histograms. 
 Note the significant spread in $\ekky$ and $\thetazh$
 that results from the cross-scale interactions with the IS turbulence. 
 
 These results indicate that strongly driven ITG turbulence can
 suppress, or stabilise, strongly driven ETG instabilities. This is in agreement
 with results from multiscale DNS
which show reduced short-wavelength heat transport in the
 presence of strongly driven ITG instabilities, see, e.g.,
 \cite{howard2016enhanced,howard2016comparison,maeyama2015cross}.
 In the following sections we explore the physical mechanisms
 that result in the stabilisation of the ETG instability
 in the $\massrt \rightarrow 0$ limit.
 
\begin{figure}
\begin {subfigure} {0.495\textwidth}
\begin{center}
\includegraphics[clip, trim=0cm 1.2cm 0cm 0.5cm, width=1.0\textwidth]{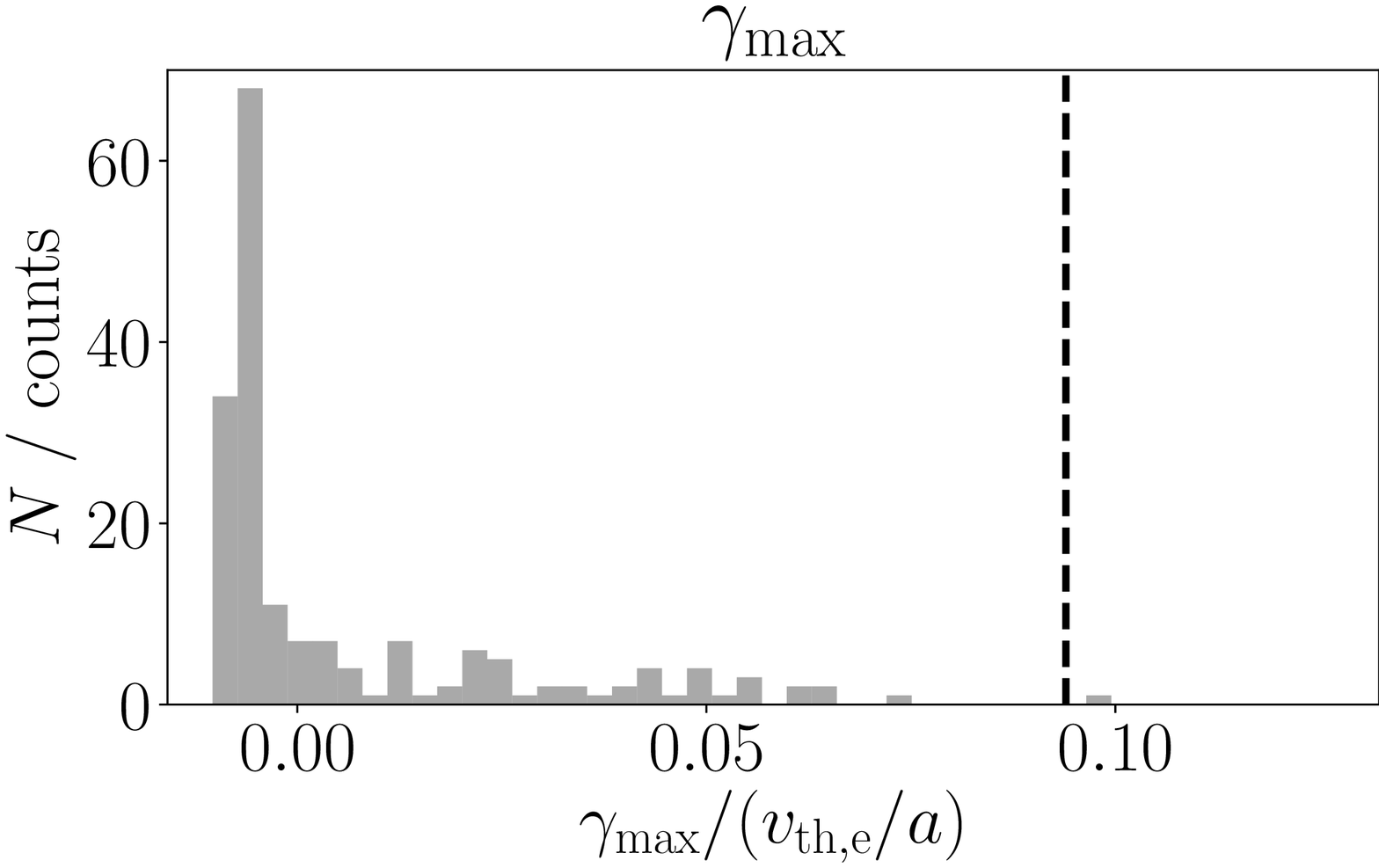}
\caption{} \label{fig:etg_growthmaxhist}
\end{center}
\end{subfigure}
\begin {subfigure} {0.495\textwidth}
\begin{center}
\includegraphics[clip, trim=0cm 1.3cm 0cm 0.5cm, width=1.0\textwidth]{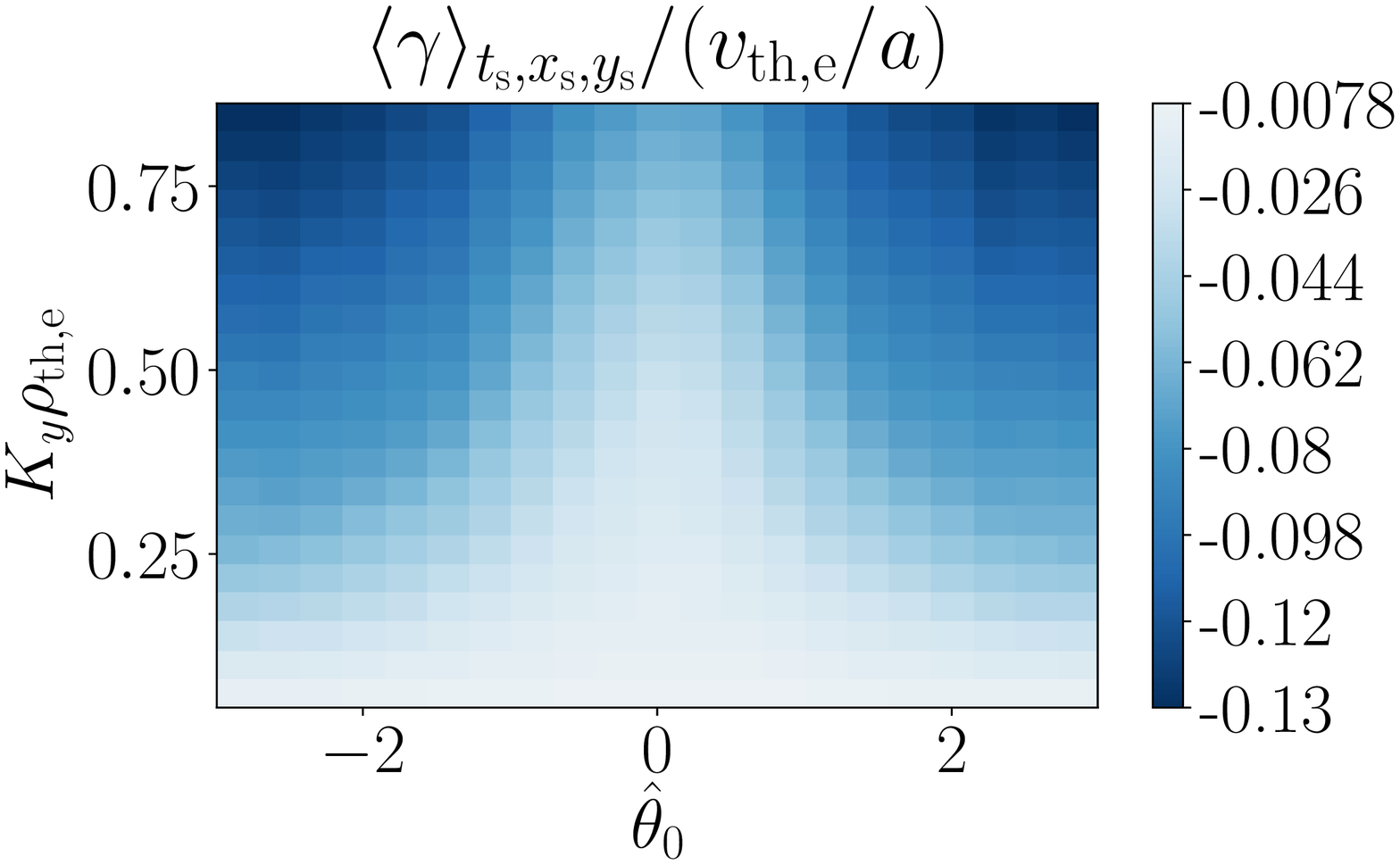}
\caption{} \label{fig:etg_fullxscale_b}
\end{center}
\end{subfigure}
\caption{ (a) The maximum ETG growth rate $\growthmax$ computed including the cross-scale terms
 $\ivee\cdot\nblf\egge{}$ and  $\evee\cdot\nbls\igge{}$
 in equation \refeq{equation:electronelectronrealgg}.
 We see that the typical effect of cross-scale interaction for these
 parameters is to stabilise the ETG mode. The dashed vertical line indicates the value of $\growthmax$ 
for the ETG mode in the absence of cross-scale interaction. 
(b) The average ETG growth rate $\avgamma$ as a function of $(\ekky,\thetazh)$.}
\label{fig:etg_fullxscale}
\end{figure}

\begin{figure}
\begin {subfigure} {0.495\textwidth}
\begin{center}
\includegraphics[clip, trim=0cm 1.2cm 0cm 0.5cm, width=1.0\textwidth]{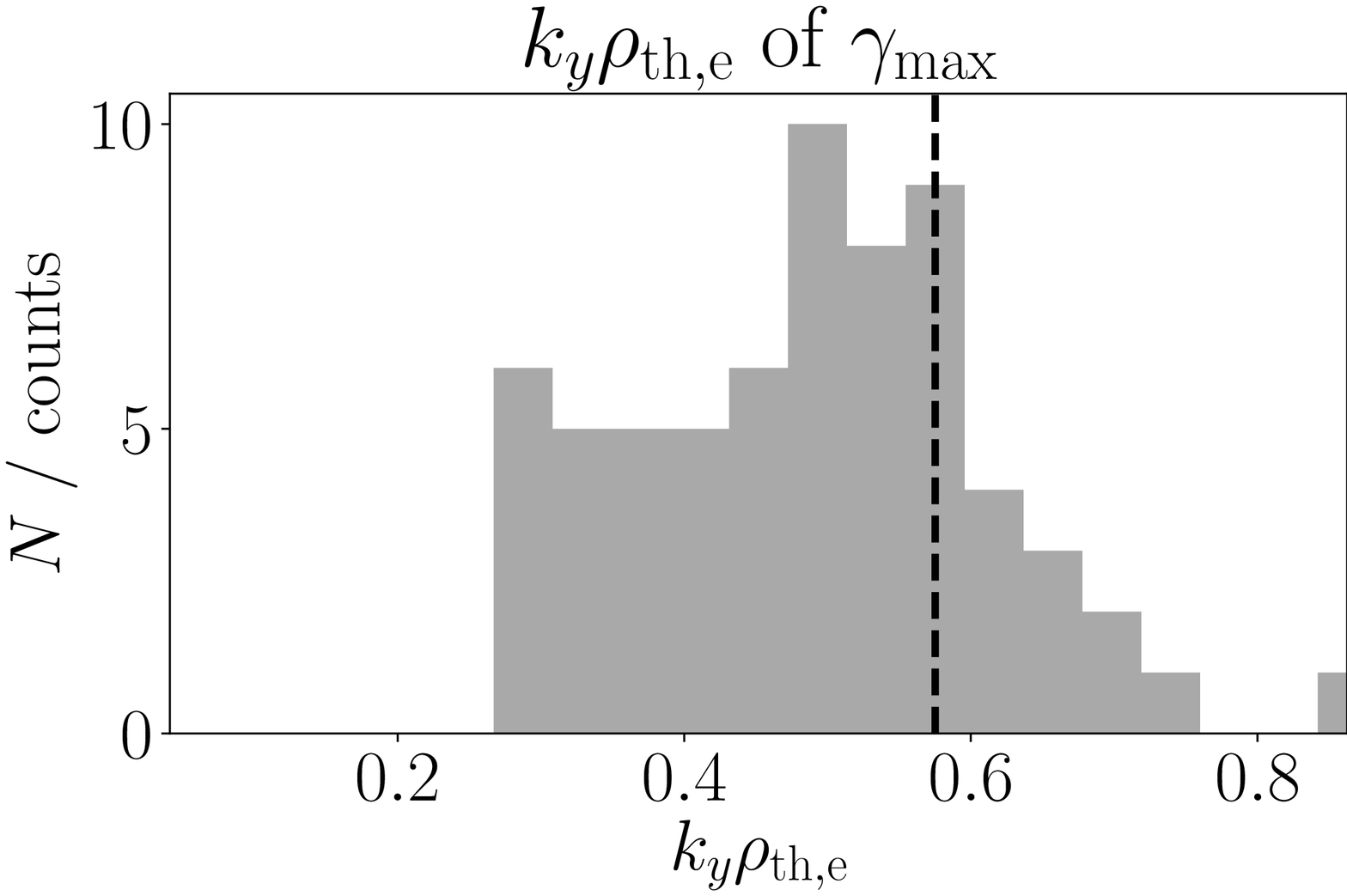}
\caption{} \label{fig:etg_ky}
\end{center}
\end{subfigure}
\begin {subfigure} {0.495\textwidth}
\begin{center}
\includegraphics[clip, trim=0cm 1.2cm 0cm 0.5cm, width=1.0\textwidth]{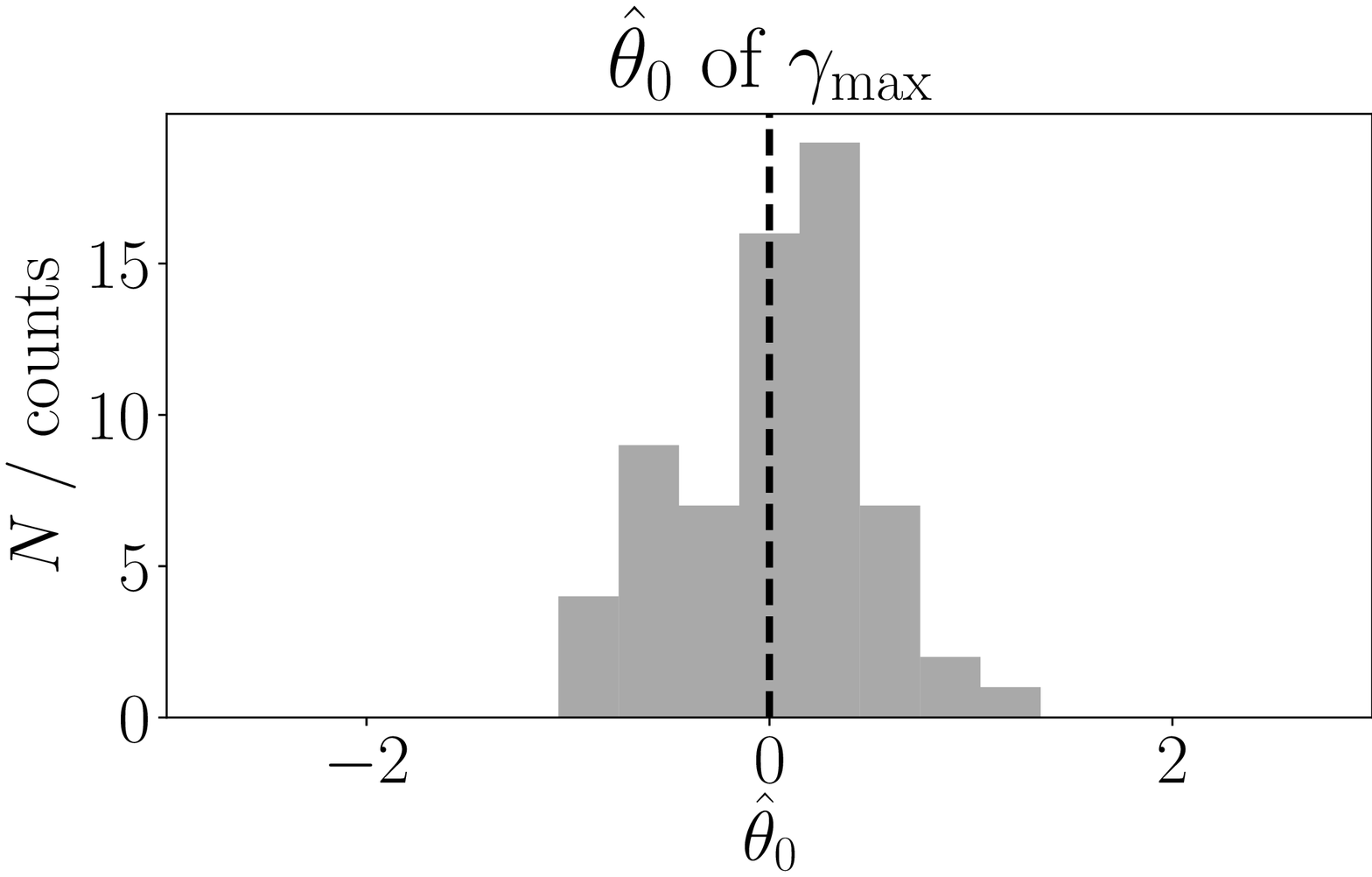}
\caption{} \label{fig:etg_theta0}
\end{center}
\end{subfigure}
\caption{ (a) The $\ekky$ of the fastest growing ETG mode.
(b) The $\thetazh$ of the fastest growing ETG mode.
The dashed lines indicate the $\ekky$ and $\thetazh$ of the ETG mode in
 the absence of IS turbulence. Only modes with $\growthmax > 0$ are included in the figures.}
\label{fig:etg_theta0ky}
\end{figure}

 \section {Physical Mechanisms and Interpretation} \label{sec:Interpretation}

The cross-scale terms in equation \refeq{equation:electronelectronrealgg},
 $\evee \cdot \nbls \igge{}$ and $\ivee \cdot \nblf \egge{}$ have
 intuitive physical interpretations.
 The gradient $\nbls \igge{}$ modifies the local gradients of the
 background Maxwellian distribution function $\nbl \eqlbe$, and
 thus modifies the instability drive.
 The drift $\ivee$ is a cross-field flow that 
 advects ES fluctuations and that can vary strongly
 in the parallel-to-the-field direction.
 This parallel-to-the-field variation introduces a new parallel length scale
and means that $\ivee$ cannot simply be removed
 from the equations by changing to a rotating or boosted frame.
 The most unstable toroidal instabilities typically have a parallel
 wave number $\kpara$ that is set by the connection length $\saffac \rmaj$,
 i.e. $\kpara \saffac \rmaj \sim 1$, cf. \cite{Parisi_arXiv_2020a}. 
 Parallel-to-the-field shearing of the ETG fluctuations
 acts to increase $\kpara$ by imposing a parallel length scale such that 
 $\kpara \saffac \rmaj \gtrsim 1$, and hence stabilises toroidal modes. 
 We illustrate the physical picture for
parallel-to-the-field shearing in figure \ref{fig:etg_shearcartoon}.
Figure \ref{fig:etg_beforeshear} depicts the field-aligned structure
 of the toroidal ETG mode in the absence of parallel shearing.
 Figure \ref{fig:etg_aftershear} shows the result of
 imposing a cross-field flow that varies
 along the magnetic field line: the $\kpara$ of the ETG mode
 is now set by the parallel-to-the-field structure of $\ivee$, and not by the
 connection length $\saffac \rmaj$.
 The drift $\ivee$ varies strongly in the direction of $\bu$ because
 at a specific $\radials$ and $\binormals$ the IS potential
 $\iptl{}=\iptl{}(\lpar)$. 
 We note that we may express
 $\nbls\igge{} = \nbls \ihhe{} + \nbls(\charge \iptl{}/\tempe)\eqlbe$,
  with $\ihhe{}$ the nonadiabatic response of electrons at IS.
  In the orbit-averaged model $\ihhe{}$ is a constant in $\lpar$, at fixed $(\energy,\pitch)$,
  and is zero in the passing piece of the velocity space for modes
  which oscillate in $\binormal$ (nonzonal modes) \cite{hardmanpaper1}.
  Despite this, $\igge{}$ can vary strongly in the direction of $\bu$ because
  of the contribution from the adiabatic response $\charge\iptl{}\eqlbe/\tempe$. 
 Hence, $\nbls\igge{}$ contains a contribution to the density gradient $\nbls(\charge \iptl{}/\tempe)\eqlbe$
 that has the effect that the total modified drive of instability $\nbls\igge{} + \nbl\eqlbe$
 drives the mode nonuniformly in $\lpar$:
 this effect can also impose an increased $\kpara$ on the mode. 
 We note that the contribution to $\nbls\igge{}$ from the nonadiabatic electron response $\nbls\ihhe{}$
 contains modifications to the background density and temperature gradients,
 and gradients of higher-order velocity moments that may drive or suppress instabilities.
 It is worth noting that the velocity moments of $\ihhe{}$ do in general depend on $\lpar$:
 this is a consequence of 
 the fact that trapped particles can only access a limited range of $\lpar$,
 and the fact that the nonzonal,
 passing electron nonadiabatic response vanishes in the orbit-averaged model.

\begin{figure}
\begin {subfigure} {0.495\textwidth}
\begin{center}
\includegraphics[clip, trim=0cm 0cm 0cm 0cm, width=1.0\textwidth]{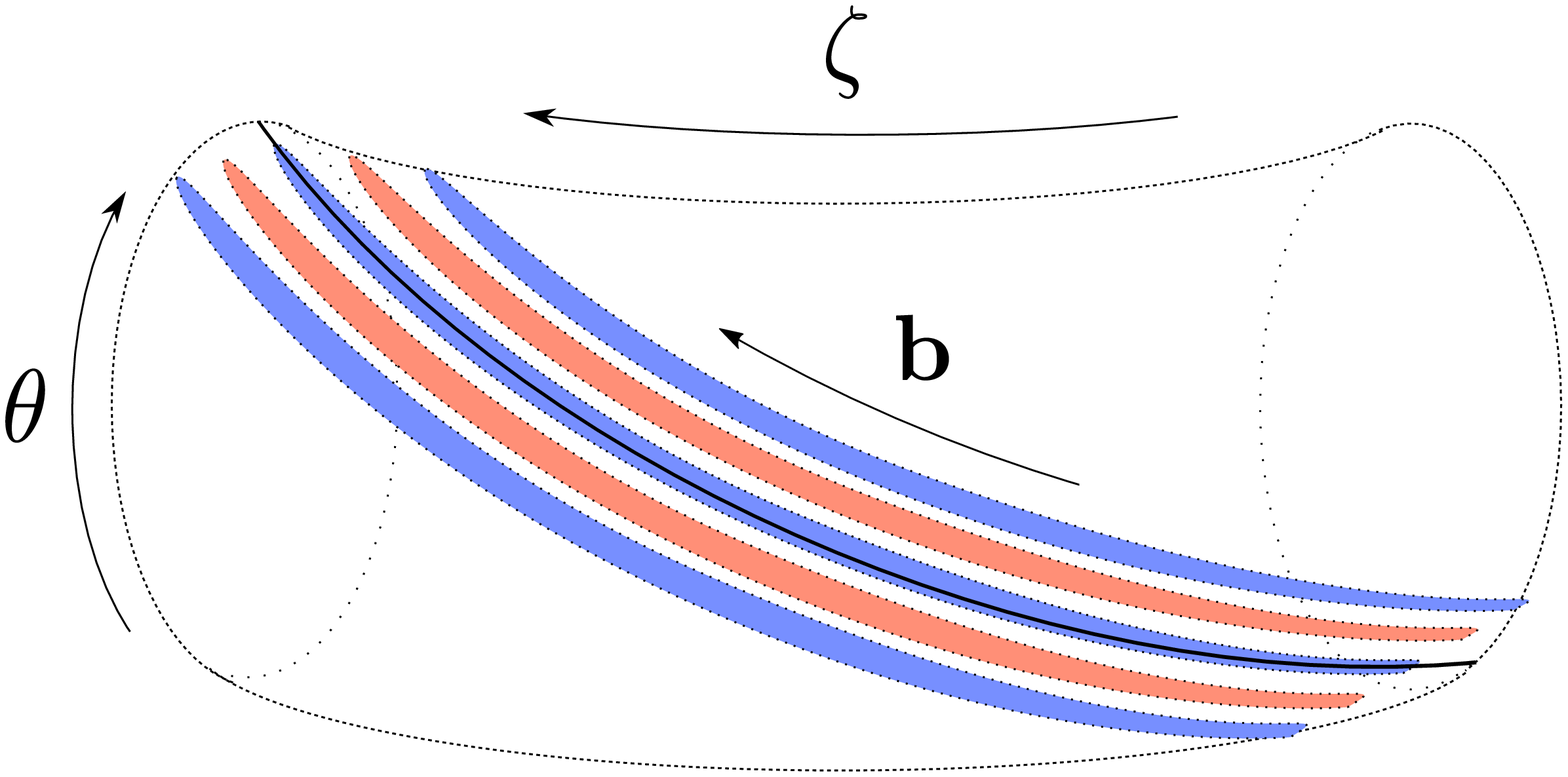}
\caption{} \label{fig:etg_beforeshear}
\end{center}
\end{subfigure}
\begin {subfigure} {0.495\textwidth}
\begin{center}
\includegraphics[clip, trim=0cm 0cm 0cm 0cm, width=1.0\textwidth]{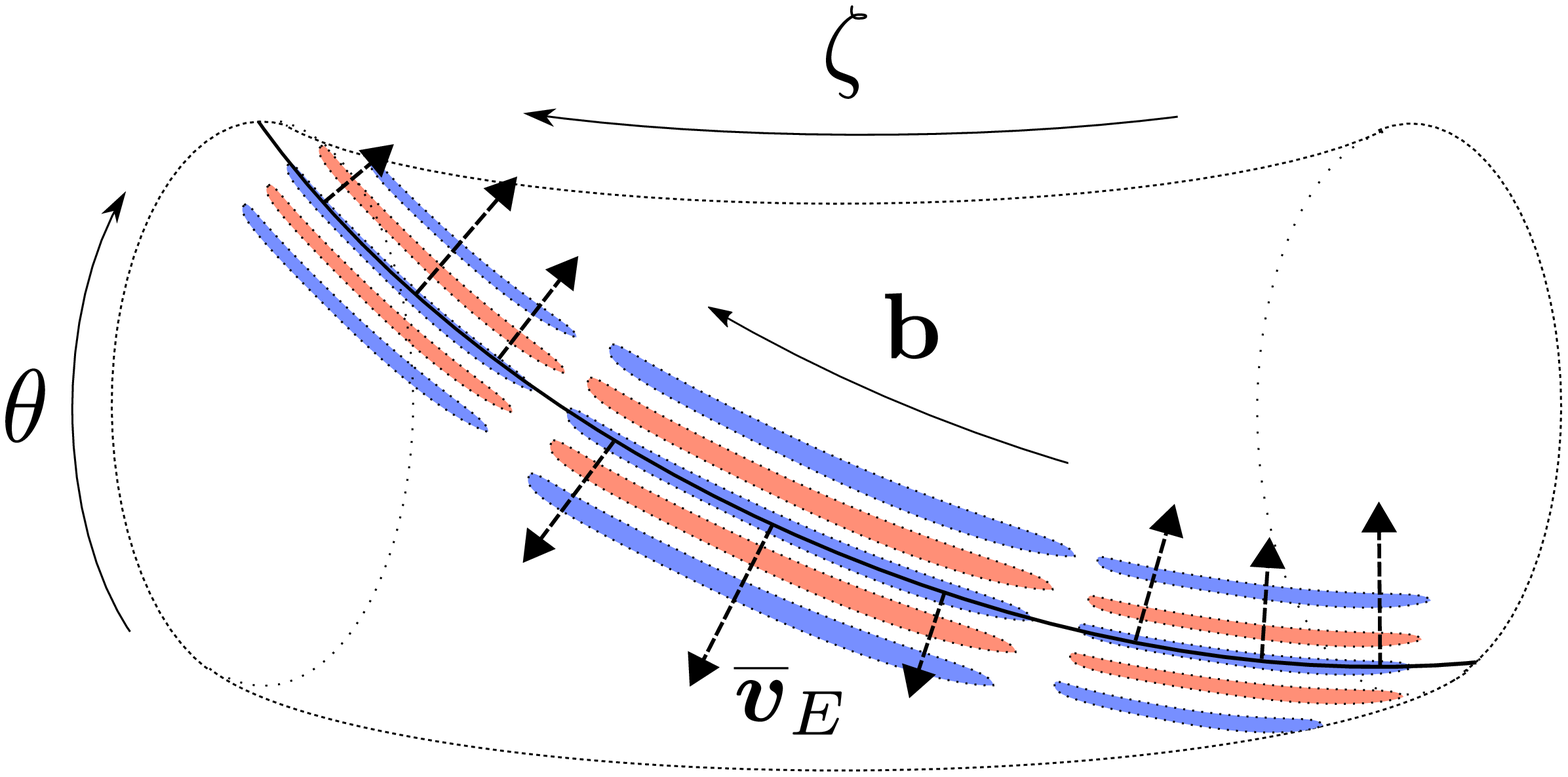}
\caption{} \label{fig:etg_aftershear}
\end{center}
\end{subfigure}
\caption{ (a) Cartoon of field-aligned contours of ETG-driven potential
 fluctuations in the absence of parallel-to-the-field shear.
(b) The same ETG-driven mode in the presence of an $\exbtext$ drift 
which varies in the direction of $\bu$ on a scale shorter than the connection length 
 by a factor of order unity:
 the parallel-to-the-field scale of the mode is shortened.}
\label{fig:etg_shearcartoon}
\end{figure}

 \begin{figure}
\begin {subfigure} {0.495\textwidth}
\begin{center}
\includegraphics[clip, trim=0cm 1.2cm 0cm 0.5cm, width=1.0\textwidth]{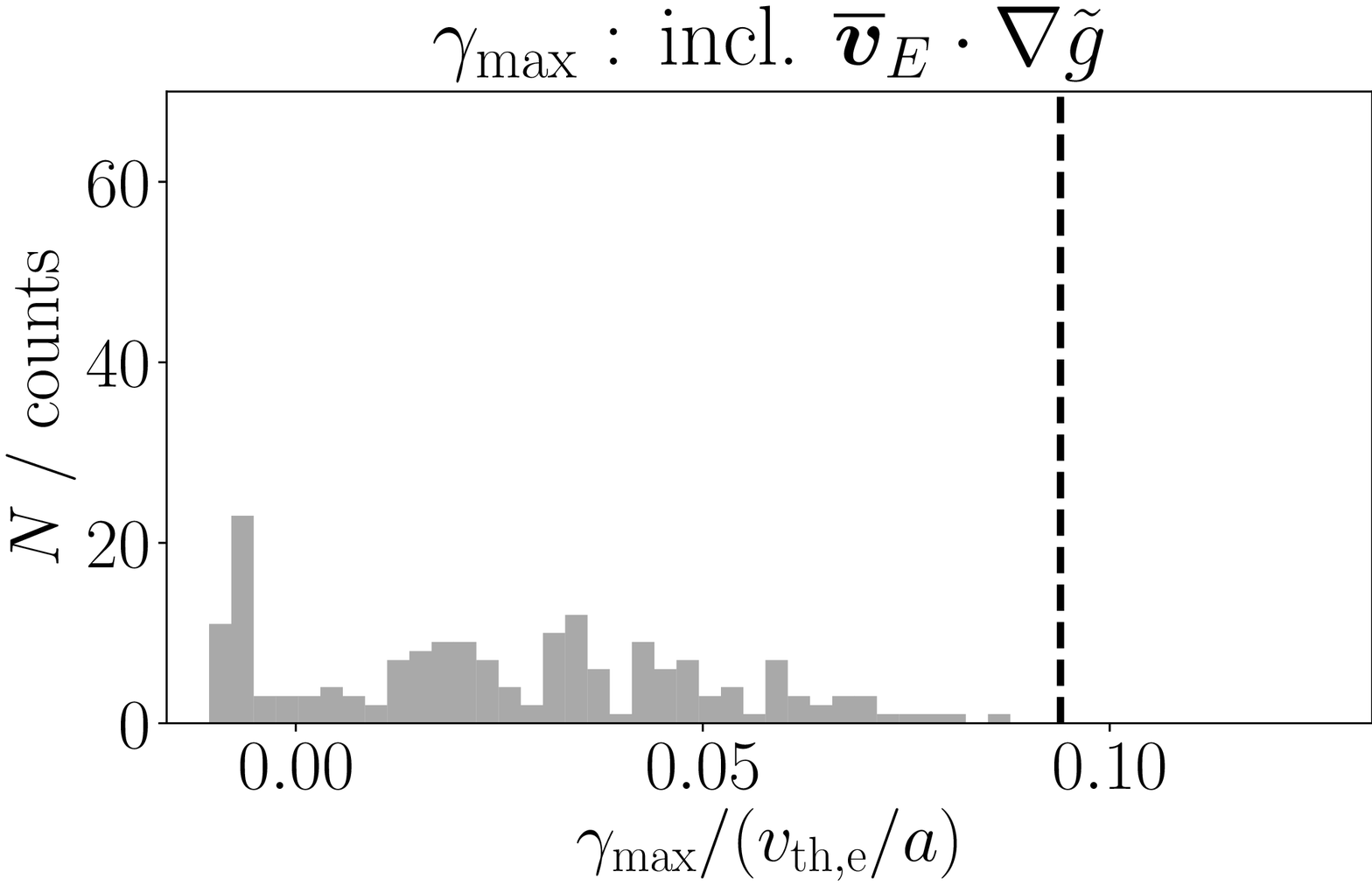}
\caption{} \label{fig:etg_shearxscale}
\end{center}
\end{subfigure}
\begin {subfigure} {0.495\textwidth}
\begin{center}
\includegraphics[clip, trim=0cm 1.2cm 0cm 0.5cm, width=1.0\textwidth]{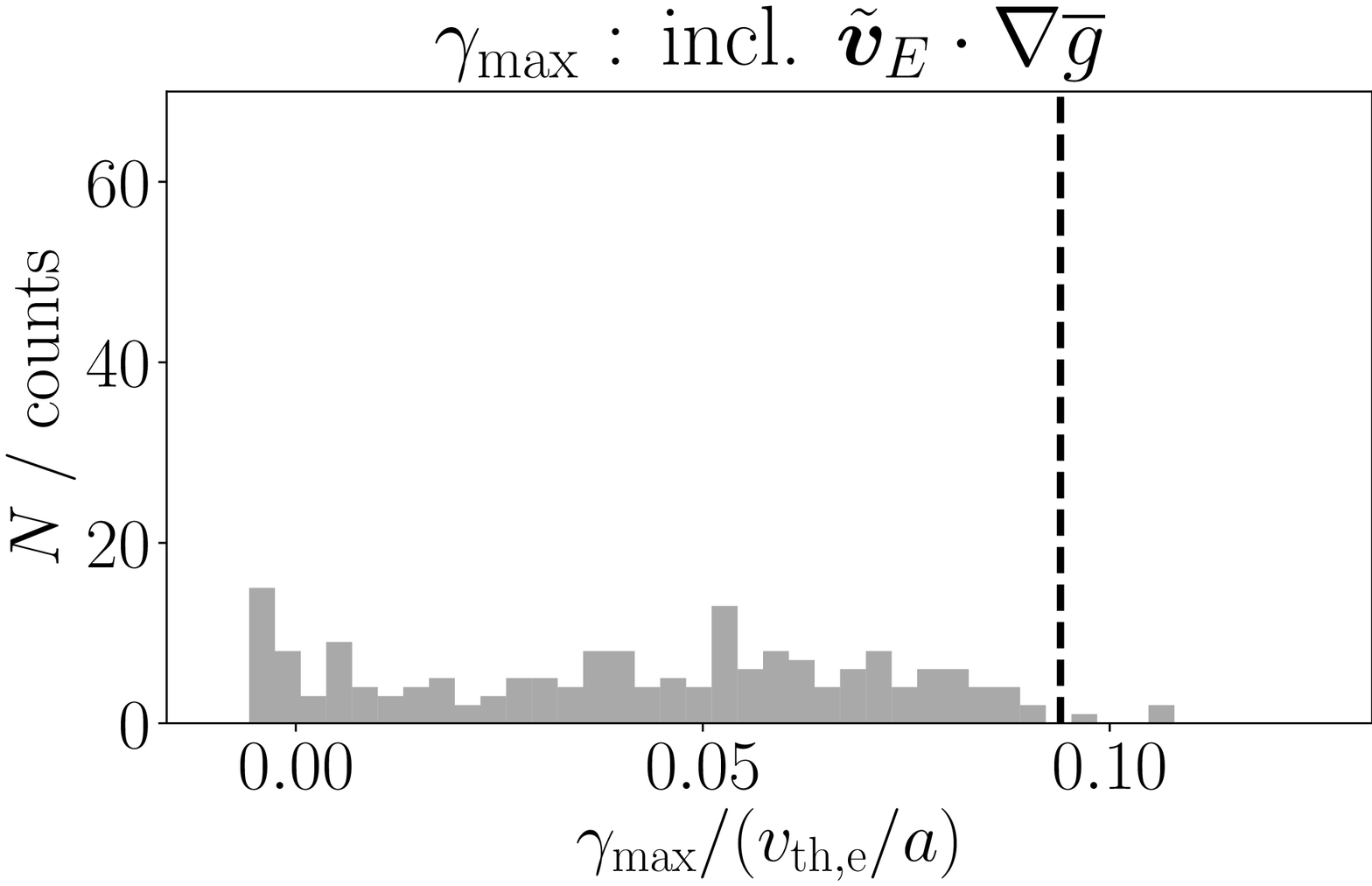}
\caption{} \label{fig:etg_gdfnxscale}
\end{center}
\end{subfigure}
\begin {subfigure} {0.495\textwidth}
\begin{center}
\includegraphics[clip, trim=0cm 1.2cm 0cm 0.5cm, width=1.0\textwidth]{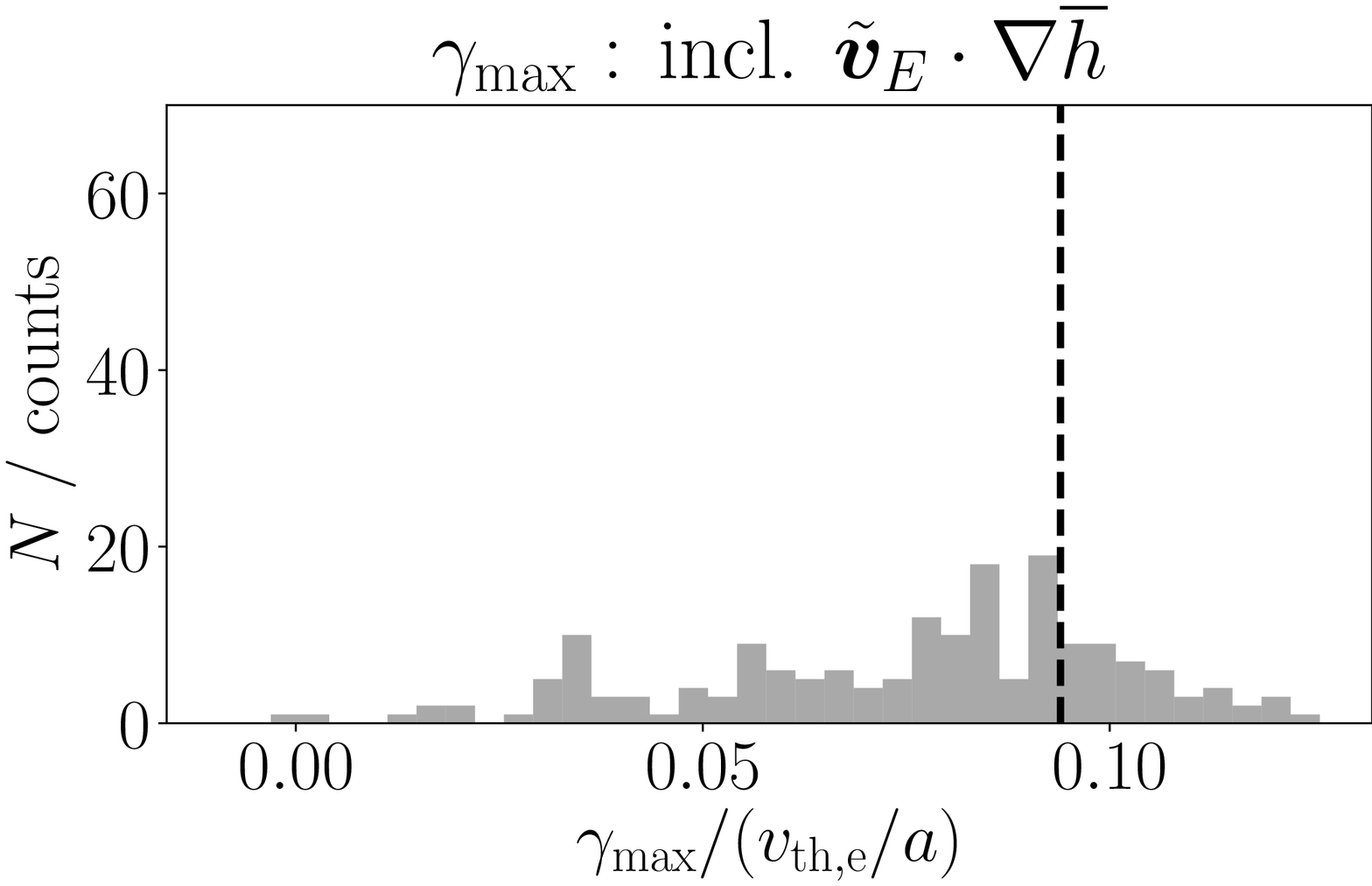}
\caption{} \label{fig:etg_hdfnxscale}
\end{center}
\end{subfigure}
\begin {subfigure} {0.495\textwidth}
\begin{center}
\includegraphics[clip, trim=0cm 1.2cm 0cm 0.5cm, width=1.0\textwidth]{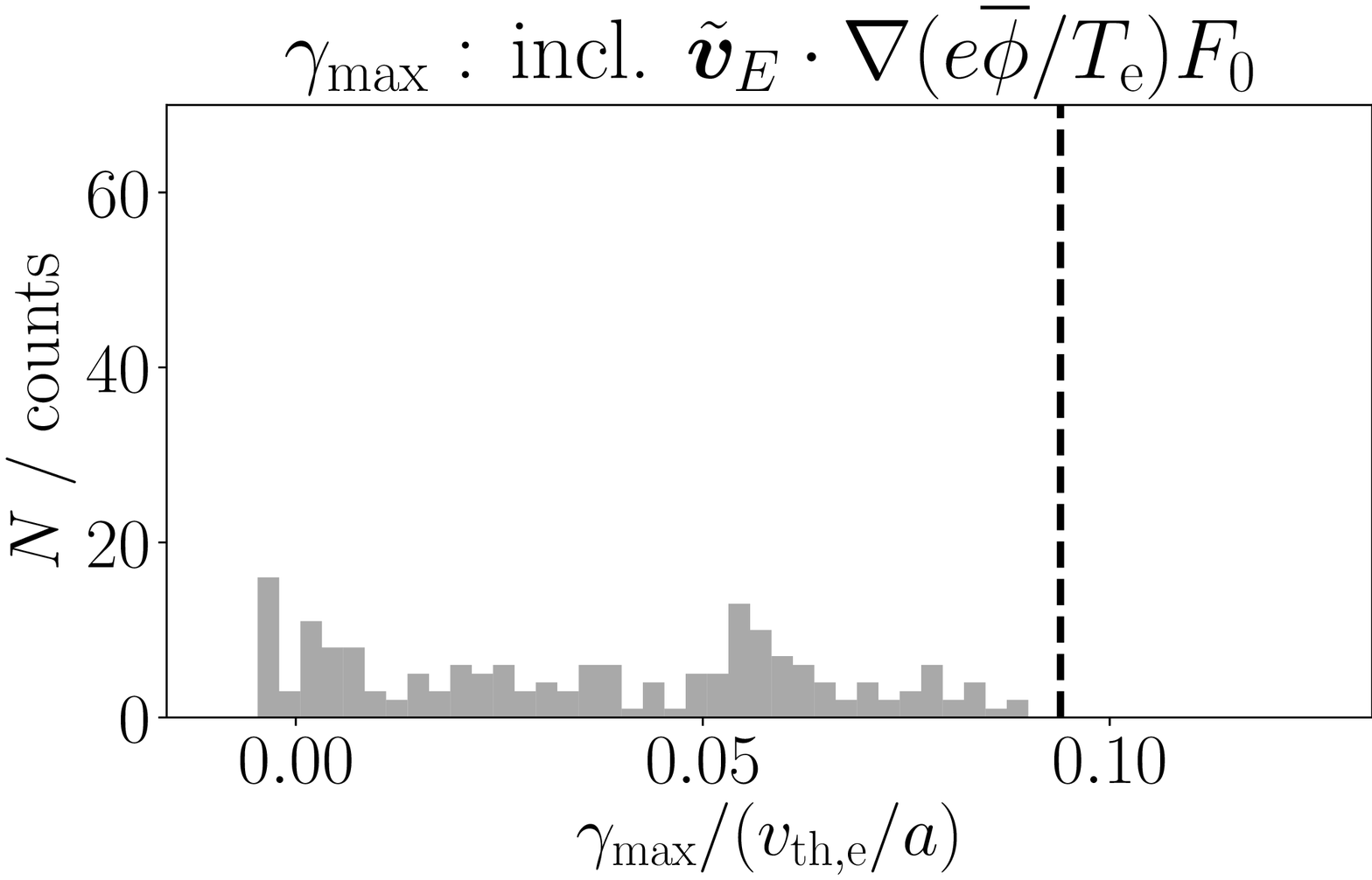}
\caption{} \label{fig:etg_phidfnxscale}
\end{center}
\end{subfigure}

\caption{Histograms of the maximum ETG growth rate $\growthmax $ computed
 using equation \ref{equation:electronelectronrealgg},
but including only the cross-scale terms indicated. The dashed vertical line
 indicates the maximum ETG growth rate in the absence of cross-scale coupling. }
\label{fig:etg_testxscale}
\end{figure}

To shed light on the physical mechanisms
 underlying the result shown in figure \ref{fig:etg_fullxscale},
 we separately consider the effect of including $\nbls \igge{}$ and $\ivee$.
 In figure \ref{fig:etg_shearxscale} we
 show the histogram of the maximum ETG growth rate $\growthmax$ 
computed including only the cross scale term $\ivee\cdot\nblf\egge{}$
 (setting $\nbls\igge{}=0$). In figure \ref{fig:etg_gdfnxscale} we
 show the histogram of  $\growthmax$ 
computed including only the cross scale term $\evee\cdot\nbls\igge{}$
 (setting $\ivee{}=0$).
 We see by comparing figures \ref {fig:etg_growthmaxhist}, 
 \ref {fig:etg_shearxscale}, and \ref{fig:etg_gdfnxscale} that both
 the effect of parallel-to-the-field shearing and modifications
 to the background gradients
are important in the ETG stabilisation shown in figure \ref {fig:etg_fullxscale}.
  From figure \ref{fig:etg_shearxscale} we see that the effect of $\ivee$
  appears to be
 uniformly stabilising; i.e., there are no instances where $\growthmax$ 
becomes larger than the single-scale ETG growth rate. However, in
 figure \ref{fig:etg_gdfnxscale} we see that the effect of
 $\nbls \igge{}$ almost always
 stabilises the ETG mode, but in rare instances can make the ETG mode more
 unstable.
 In the following sections we study these cross-scale physics mechanisms
 in more detail and provide qualitative explanations for these results.
 
 We note that our decomposition of the cross-scale terms,
 while physically motivated, is not unique. In particular,
 we recall that we
 can further decompose
 $\nbls \igge{} = \nbls \ihhe{} + \nbls(\charge \iptl{}/\tempe)\eqlbe$. 
 In figures \ref{fig:etg_hdfnxscale} and \ref{fig:etg_phidfnxscale} we
 show the cross-scale effects of $\nbls \ihhe{}$ and
 $\nbls(\charge \iptl{}/\tempe)\eqlbe$ on the ETG mode separately.
 The piece due to the nonadiabatic response $\nbls \ihhe{}$ 
 can both
 drive and suppress the ETG mode, whereas the piece due to the
 adiabatic response $\nbls(\charge \iptl{}/\tempe)\eqlbe$
 appears to only suppress the ETG mode.
 We give a qualitative explanation for this observation in section \ref{sec:fptp}.
 
 \section {Evidence for parallel-to-the-field shearing} \label{sec:exb}
 In this section we show that the physical picture given
 in figure \ref{fig:etg_shearcartoon} is consistent with the simulation
 results presented in section \ref{sec:Numerical}: the IS
 turbulence indeed shortens the parallel-to-the-field scale of the ETG modes.
 To demonstrate this 
 we first consider the structure of the cross-scale terms and the structure
 of the ETG eigenmode in the ballooning coordinate, labelled by $\lpar$.
 The fastest-growing ETG mode from a single-scale simulation
 with CBC parameters occurs at $\kky\gyrde = 0.58$ and $\thetazh = 0$. 
 Figure \ref {fig:etg_eigmode} compares the potential eigenmode for the $(\kky\gyrde = 0.58,\thetazh = 0)$ mode
 from a single-scale simulation, to the $(\kky\gyrde = 0.58,\thetazh = 0)$
 mode obtained at a specific example space-time location in the IS turbulence.
 We see that the cross-scale interaction introduces oscillatory features into the
 potential eigenmode. This is a result of the variation of $\ivee$ and
 $\nbls \igge{} = \nbls (\charge \iptl{}/\tempe) \eqlbe + \nbls \ihhe{}$ along the magnetic field line.
 \begin{figure}
\begin {subfigure} {0.495\textwidth}
\begin{center}
\includegraphics[clip, trim=0cm 1.2cm 0cm 0cm, width=1.0\textwidth]{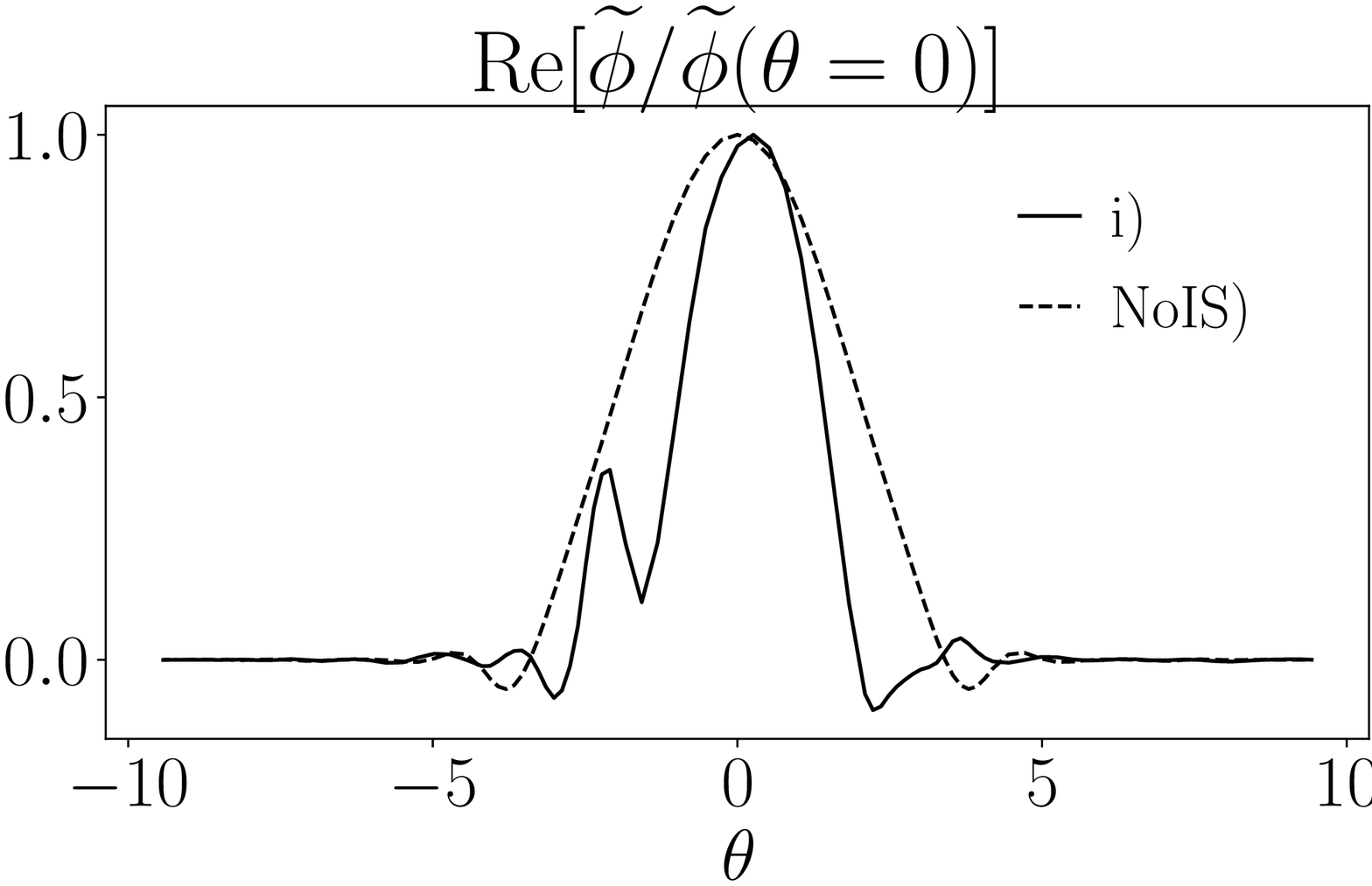}
\caption{} \label{fig:etg_eigmode}
\end{center}
\end{subfigure}
\begin {subfigure} {0.495\textwidth}
\begin{center}
\includegraphics[clip, trim=0cm 1.2cm 0cm 0cm, width=1.0\textwidth]{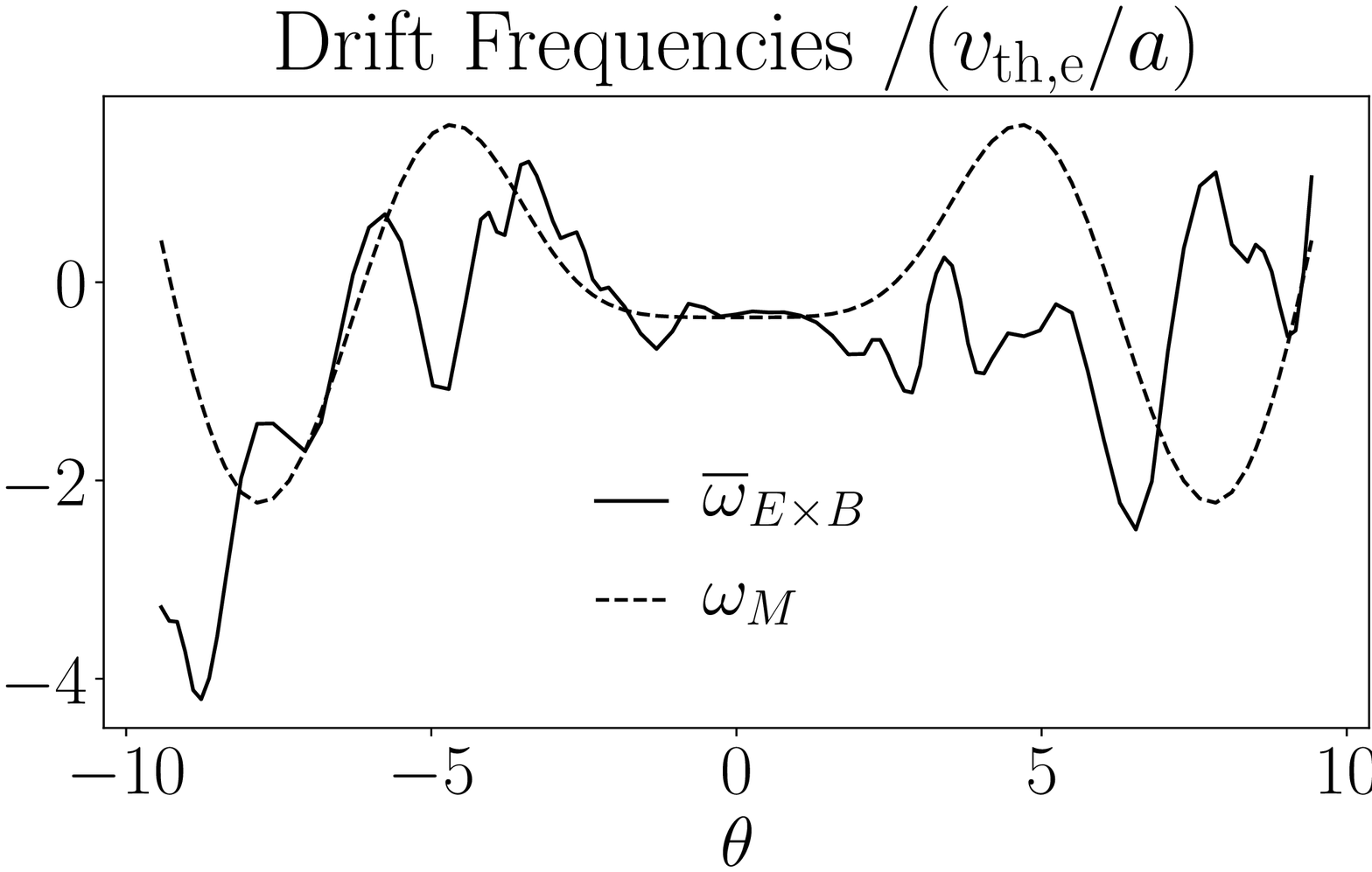}
\caption{} \label{fig:etg_wdriftwexb}
\end{center}
\end{subfigure}
\begin {subfigure} {0.495\textwidth}
\begin{center}
\includegraphics[clip, trim=0cm 1.2cm 0cm 0cm, width=1.0\textwidth]{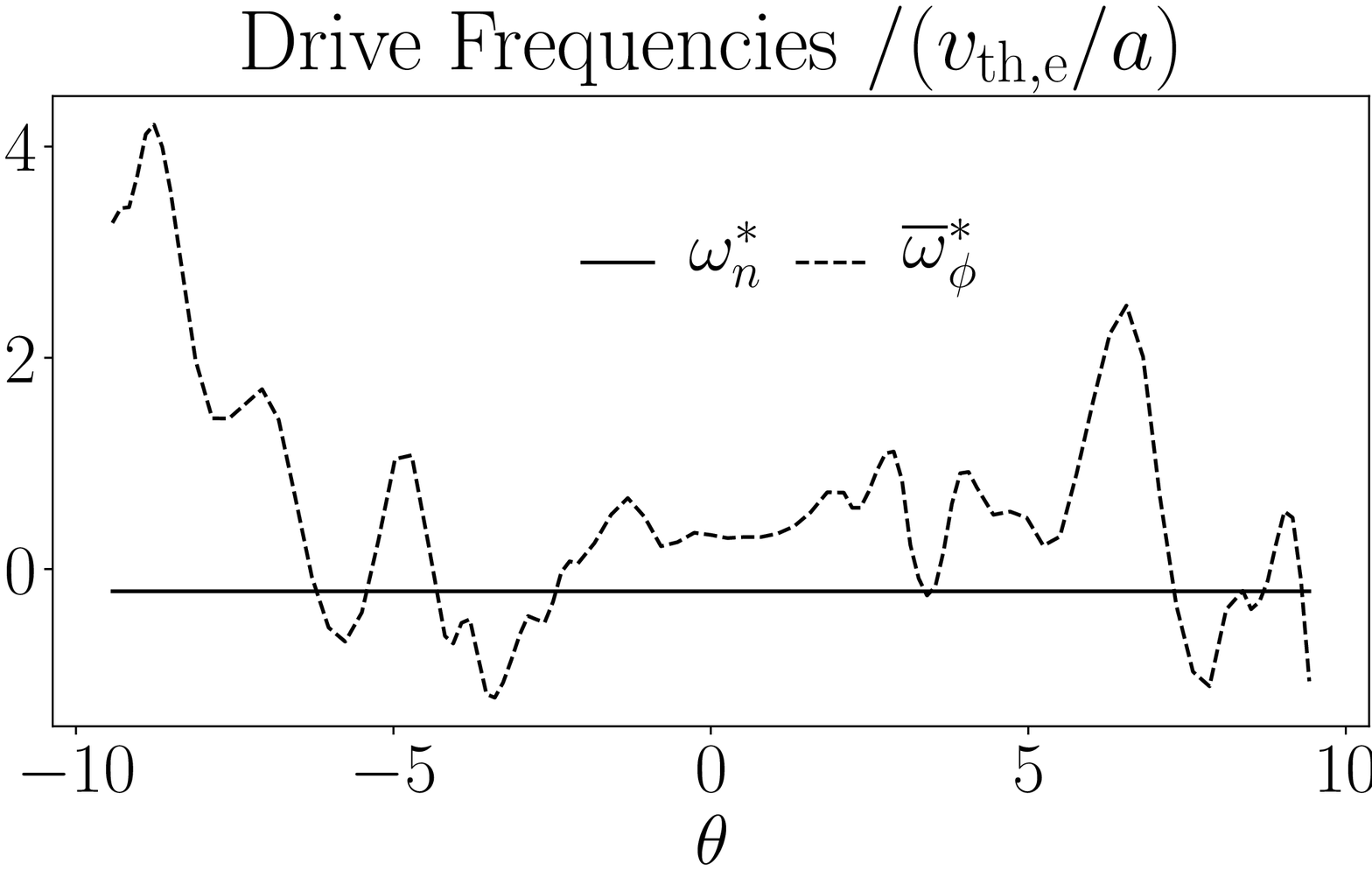}
\caption{} \label{fig:etg_wstarphi}
\end{center}
\end{subfigure}
\begin {subfigure} {0.495\textwidth}
\begin{center}
\includegraphics[clip, trim=0cm 1.2cm 0cm 0cm, width=1.0\textwidth]{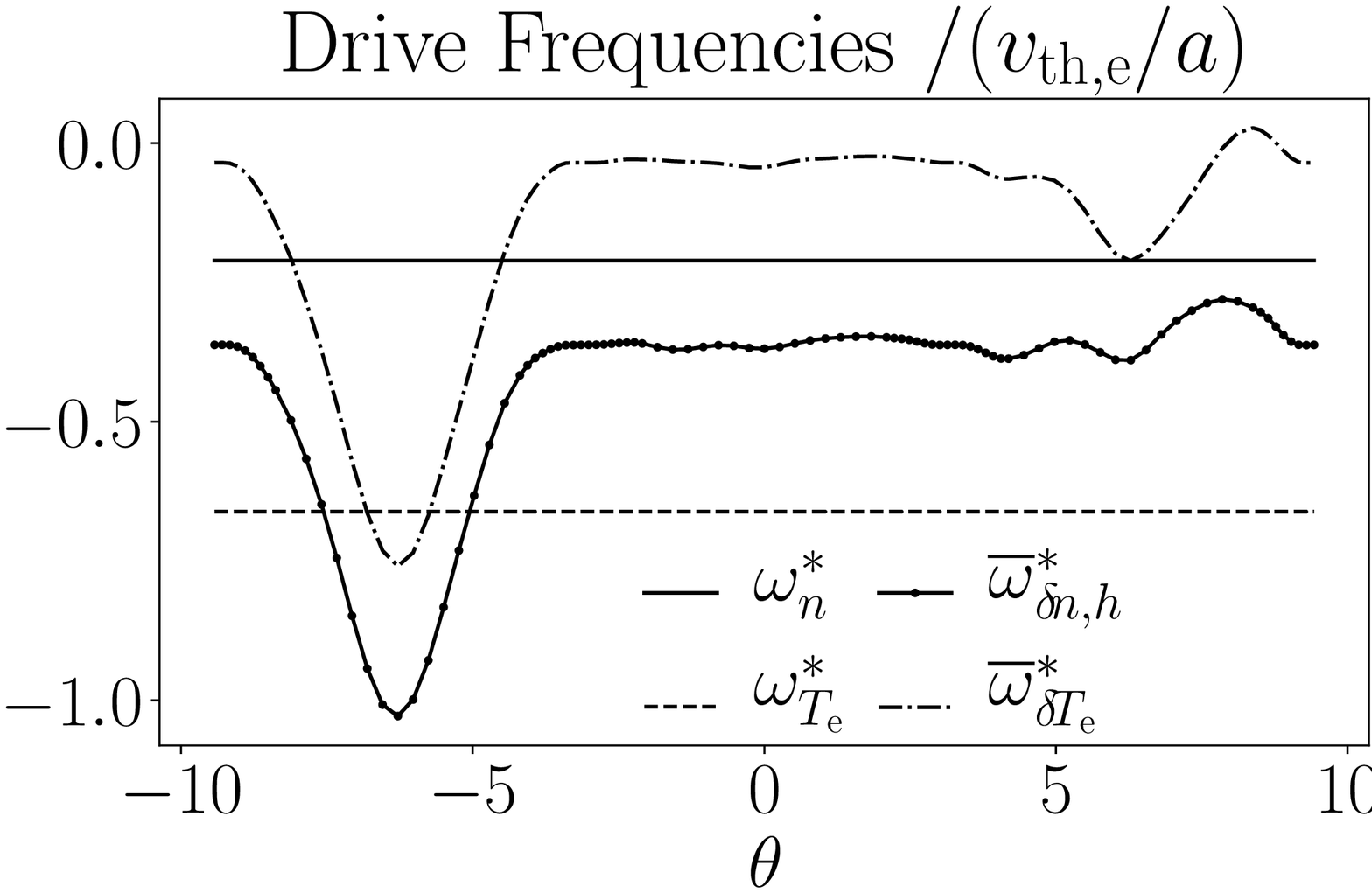}
\caption{} \label{fig:etg_wstart}
\end{center}
\end{subfigure}
\caption{ (a) The real part of the normalised potential eigenmode for the fastest-growing
 ETG mode for i) a specific instance of IS gradients and NoIS) the eigenmode in the 
 absence of cross-scale coupling.
 The normalised growth rates are $\growth / (\vthere/\lscal) = $ 0.045 and 0.094, respectively.
 For simplicity, we chose to compare a case where the fastest-growing mode in the prescence of IS 
 turbulence appears at $\ekky \gyrde = 0.58$ and $\thetazh = 0.0$. 
(b) The thermal magnetic drift frequency $\wm $ compared to the IS $\exbtext$ drift frequency $\iwe $.
(c) The drive frequencies due to equilibrium density gradients $\wstarn$,
 and due to gradients in the adiabatic part of the IS, electron distribution function $\wstarphi$.
 (d) The drive frequencies due to equilibrium temperature gradients $\wstart$,
 and gradients of temperature and density in the nonadiabatic part of the
 IS, electron distribution function, $\wstardt$ and $\wstardn$, respectively. }
\label{fig:etg_coeff_examples}
\end{figure}
 To illustrate this, in figure \ref{fig:etg_wdriftwexb}
  we show  the imposed IS $\exbtext$ drift frequency
  \beqn \iwe = 
  \frac{\ltsp \kxfac}{\bscal}\left(\ekkx \drv{\iptl{}}{\binormal}
 - \ekky \drv{\iptl{}}{\radial} \right), \eeqn
 with
 $\ekkx$ the field-aligned radial wave number
 of the ETG mode, i.e., the wavenumber
 $\ekkvec = \ekkx \nbl \radial + \ekky \nbl \binormal$, and the
 geometrical factor $\kxfac = \kxfacdef \simeq 1.01$ for CBC parameters.
 The frequency $\iwe$ is compared to
 the thermal magnetic drift frequency \beqn
  \wm = \frac{\vthere^2}{\cycfe}\ekkvec \cdot \bu \xp \left( \frac{\nbl \bmag}{\bmag} + \bu \cdot \nbl \bu \right). \eeqn 
 In figure \ref {fig:etg_wstarphi}
 we show the drive frequencies associated with
 the $\nbls (\charge \iptl{}/\tempe) \eqlbe$ piece of $\nbls \igge{}$, $\wstarphi = - \iwe$,
 compared to the background density gradient drive
 $\wstarn = - \ltsp \ekky \tempe / \charge \bscal \lln$;
 and in figure \ref {fig:etg_wstart} we show the background temperature gradient drive
 frequency $\wstart =  - \ltsp \ekky \tempe / \charge \bscal \lte$, and 
the drive frequencies due to the density $\inh$ and temperature $\ith$ moments of $\nbls \ihhe{}$,
 \beqn \wstardn = \frac{\ltsp \kxfac \tempe}{\charge \bscal }\left(\frac{\ekky}{\dens}
 \drv{\inh}{\radial} - \frac{\ekkx}{\dens}
 \drv{\inh}{\binormal} \right) \eeqn and
 \beqn \wstardt = \frac{\ltsp \kxfac \tempe }{\charge \bscal}\left(\frac{\ekky}{\tempe}
 \drv{\ith}{\radial} - \frac{\ekkx}{\tempe}
  \drv{\ith}{\binormal} \right), \eeqn
 respectively. There are two key feature to note in figures \ref{fig:etg_wdriftwexb},
 \ref{fig:etg_wstarphi}, and \ref{fig:etg_wstart}. Firstly, the frequencies that
 mediate the effects of IS turbulence
 on ETG modes are comparable in amplitude to the equilibrium drive
 and thermal magnetic frequencies. Secondly, the frequencies
 vary in ballooning $\lpar$ on the scale of the
 ETG eigenmode in \ref{fig:etg_eigmode}. In the strongly driven turbulence that we 
 consider here, we have found that
 the $\lpar$ dependence of $\iwe$ 
 cannot be
 described by a simple function with few parameters.
 In consequence, simplified quantitative modelling of the effects of
 parallel-to-the-field shearing by $\ivee$ is 
 challenging. As we discuss in section \ref{sec:fptp},
 we find that it is possible to approximate $\wstardn$ and $\wstardt$ by a constant value.

\begin{figure}
\begin{center}
\includegraphics[clip, trim=0cm 1.2cm 0cm 0.5cm, width=0.6\textwidth]{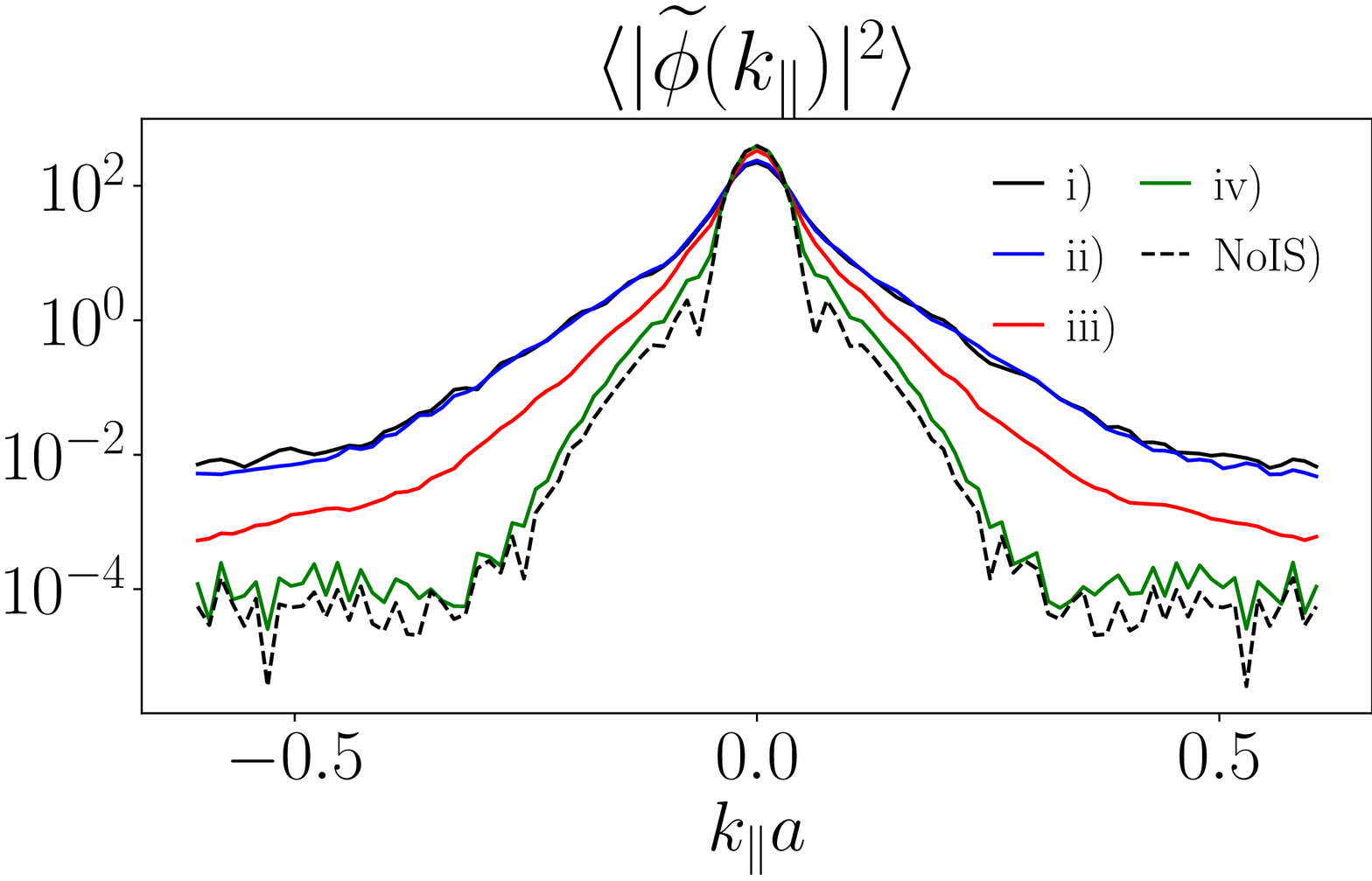}
\caption{A comparison of the power spectrum of the electrostatic
 potential $\eptl{}$ of the fastest growing ETG ballooning mode
 as a function of $\kpara$ for several cases. In case NoIS) we show the power
 spectrum of the fastest growing ballooning mode in the
 absence of cross-scale coupling. The remaining curves show the $\ispot$-averaged
 power spectrum of the fastest growing ETG mode in the case that:
 i) both cross-scale terms $\ivee \cdot \nblf \egge{}$ and $\evee \cdot \nblf \igge{}$
 are included; ii) only the shearing term $\ivee \cdot \nblf \egge{}$ 
 is included; iii) only the modification to $\nbl \eqlbe$
 from the adiabatic response of electrons is included,
 i.e., $\evee \cdot \nbls (\charge \iptl{}/\tempe) \eqlbe$; and iv)
 only the modification to $\nbl \eqlbe$
 from the nonadiabatic response of electrons is included,
 i.e., $\evee \cdot \nbls \ihhe{}$. 
 To measure the spread of the distribution of $\kpara$ we use $\sigkp$,
 defined in equation \refeq{eq:sigkp}.
 We find i) $\sigkp =  0.065$, ii) $\sigkp =  0.064$, iii) $\sigkp =  0.043$, 
 iv) $\sigkp =  0.031$, and NoIS) $\sigkp =  0.027$. } \label{fig:etg_kparcomparison}
\end{center}
\end{figure}

To assess whether or not the physical picture in figure \ref{fig:etg_shearcartoon}
is consistent with the simulation results, we calculate the
$\kpara$ spectrum of the fastest growing ETG mode at each
$\ispot$ in the sample. For a single ballooning mode, we define
the $\kpara$ spectrum to be $|\eptl{}(\kpara)|^2$, where
\beqn \eptl{}(\kpara) = \int^{\infty}_{-\infty}  \exp[\imag \kpara \arcl] \eptl{}(\arcl) d \arcl,\eeqn
cf. \cite{Parisi_arXiv_2020a}, with $\arcl = \lpar / \kpar$,
 and $\eptl{}(\arcl)$ is normalised so that the maximum value of $\eptl{}(\arcl)$ is $1$ -- 
 this maximum may occur away from $\arcl = 0$. 
We note that we define the $\lpar$ coordinate in the simulations
 such that $\kpar$ is constant in $\lpar$.
In figure \ref{fig:etg_kparcomparison} we compare 
$ |\eptl{}(\kpara)|^2  $ for the fastest growing ETG mode
 in the absence of cross-scale interaction with
the average $\kpara$ spectrum $\langle |\eptl{}(\kpara)|^2 \rangle$ of the fastest
 growing ETG modes in several calculations including various
 combinations of the cross-scale interaction terms.
 Figure \ref {fig:etg_kparcomparison}
 indicates that cross-scale interaction introduces tails in the $\kpara$ spectra.
 These tails correspond to the oscillatory structure of the eigenmode shown in
 figure \ref{fig:etg_eigmode}. To quantify the size of these tails, 
 we introduce a measure of the spread of the distribution of $\kpara$:
 \beqn \frac{\sigkp^2}{2} = \frac{\int \kpara^2 \langle |\eptl{}(\kpara)|^2 \rangle d \kpara}{
 \int \langle |\eptl{}(\kpara)|^2 \rangle d \kpara} \label{eq:sigkp}.\eeqn
 For the fastest-growing ETG mode in the absence of cross-scale interaction $\sigkp = 0.027$.
 For the average $\kpara$ spectrum of the fastest-growing ETG modes
 in the presence of all cross-scale interaction terms we find that $\sigkp = 0.065$.
 This indicates that the effect of cross-scale interaction is
 indeed to introduce larger $\kpara$ into the ETG mode. 
 Figure \ref{fig:etg_kparcomparison} indicates that the dominant
 component of the $\kpara$ tails is introduced by the parallel
 shearing term $\ivee \cdot \nblf \egge{}$.
 We also note that the term $\evee{}\cdot \nbls\ihhe{}$ introduces almost no $\kpara$ component:
 cross-scale interaction via $\evee{}\cdot \nbls \ihhe{}$ leaves the form of the ETG eigenmode unchanged.
 This is consistent with the fact that $\ihhe{}$ is constant
 in $\lpar$ at fixed $(\energy,\pitch,\sign,\radial,\binormal)$ in the parallel-orbit-averaged model.

 \section {Modifications to the drives of instability} \label{sec:fptp}
\begin{figure}
\begin {subfigure} {0.495\textwidth}
\begin{center}
\includegraphics[clip, trim=0cm 1.2cm 0cm 0.5cm, width=1.0\textwidth]{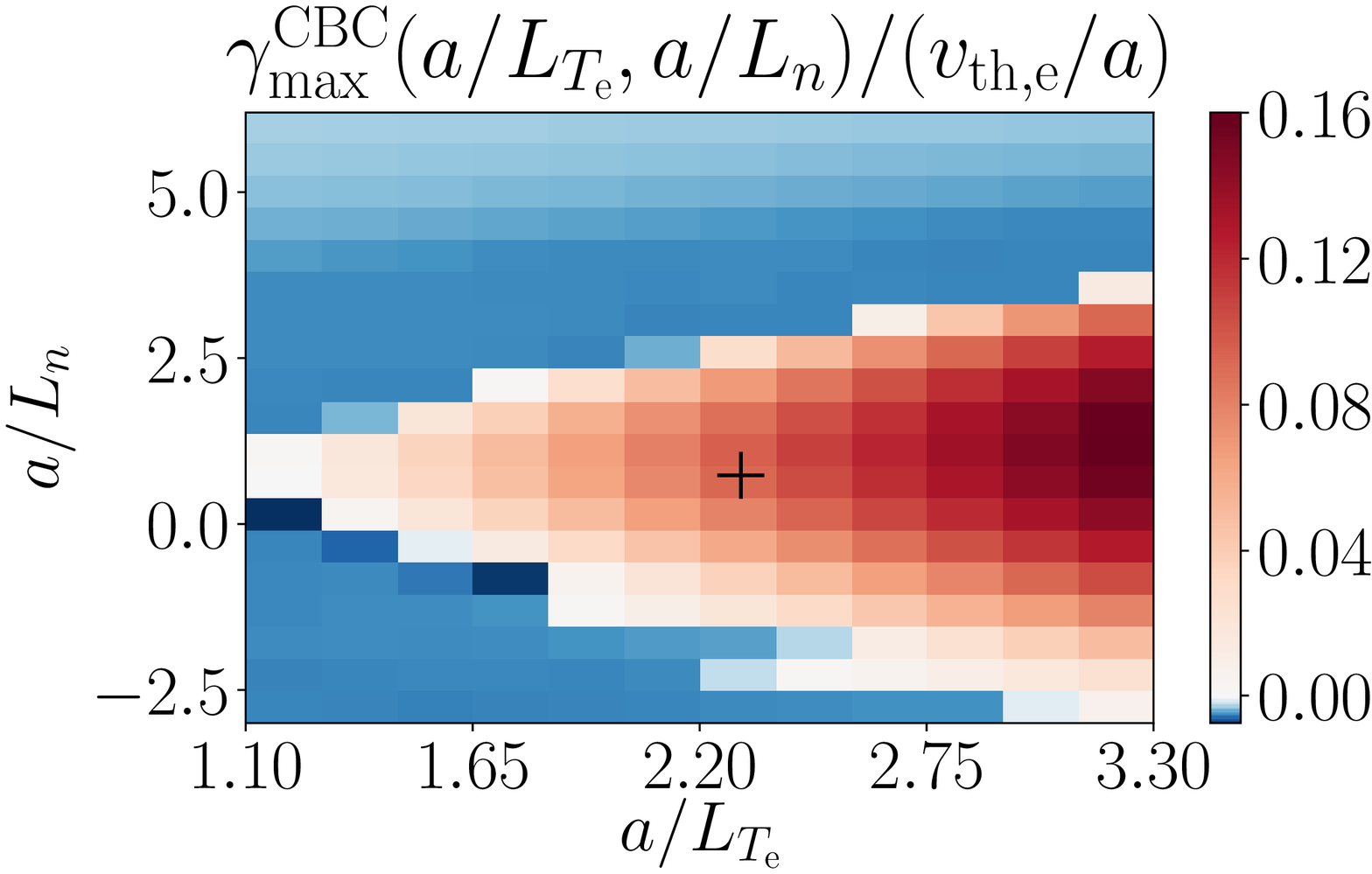}
\caption{} \label{fig:etg_fptp}
\end{center}
\end{subfigure}
\begin {subfigure} {0.495\textwidth}
\begin{center}
\includegraphics[clip, trim=0cm 1.2cm 0cm 0.5cm, width=1.0\textwidth]{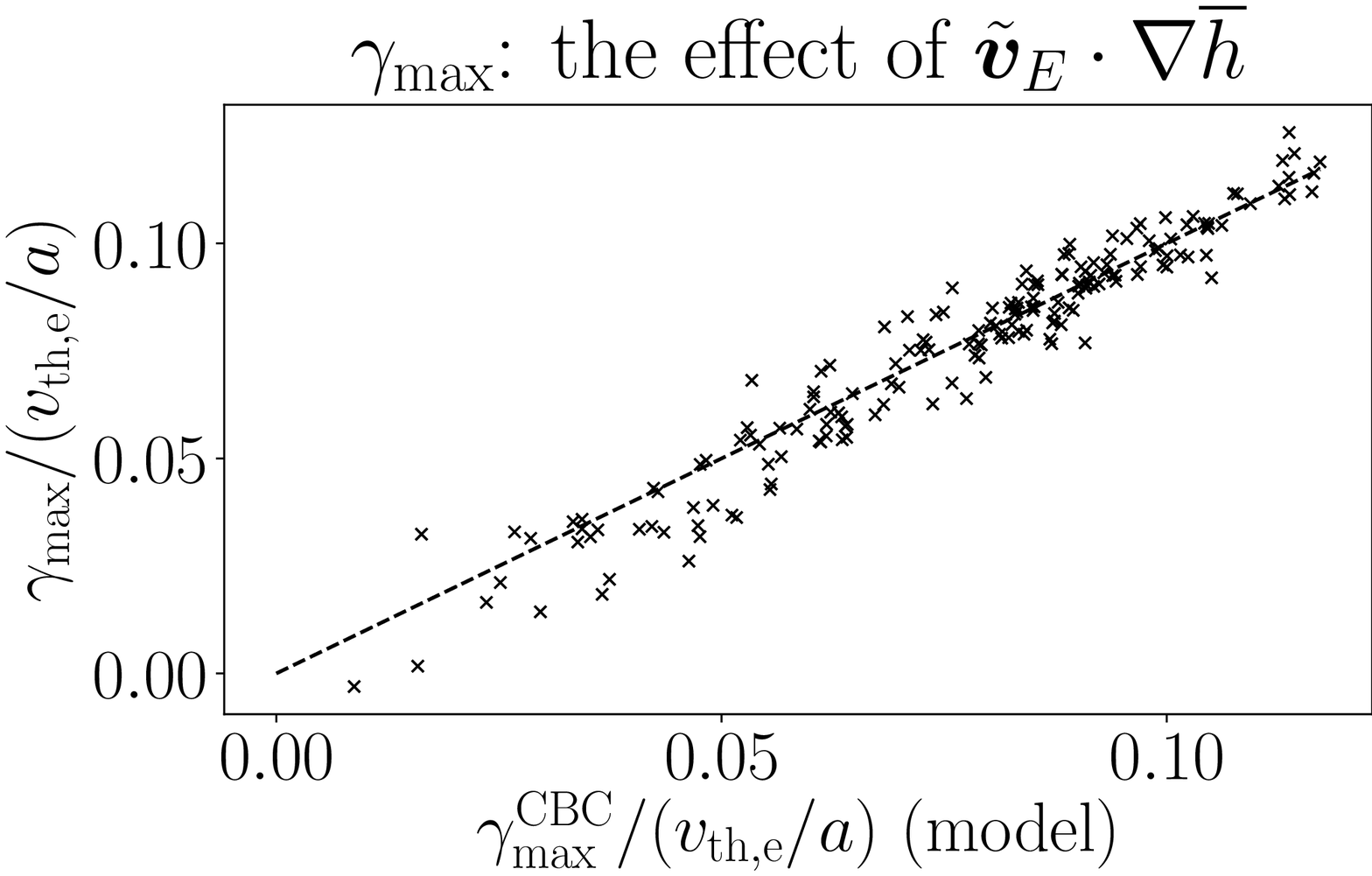}
\caption{} \label{fig:etg_dfn_fptp}
\end{center}
\end{subfigure}
\begin {subfigure} {0.495\textwidth}
\begin{center}
\includegraphics[clip, trim=0cm 1.2cm 0cm 0.5cm, width=1.0\textwidth]{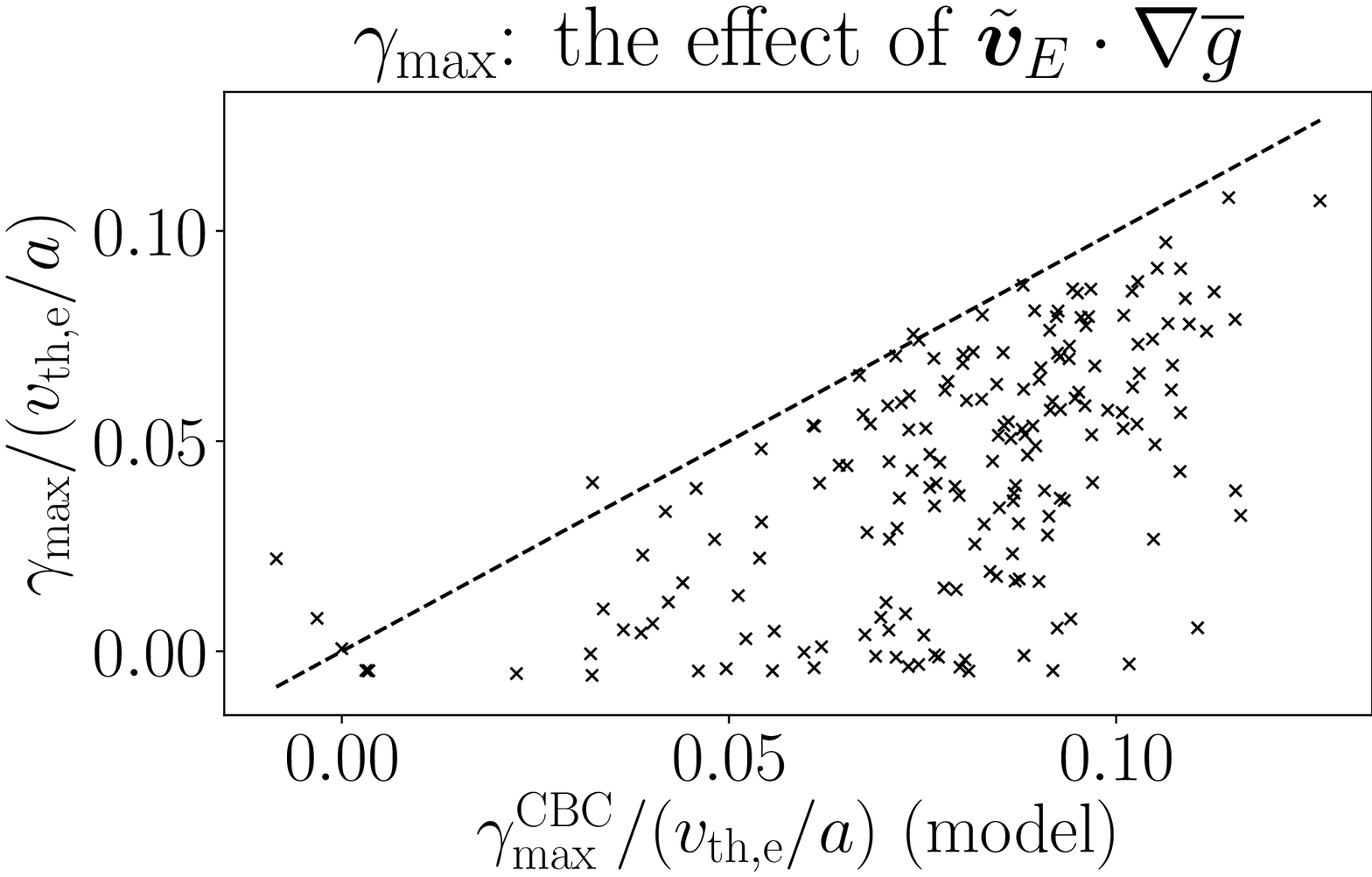}
\caption{} \label{fig:etg_gdfn_fptp}
\end{center}
\end{subfigure}
\begin {subfigure} {0.495\textwidth}
\begin{center}
\includegraphics[clip, trim=0cm 1.2cm 0cm 0cm, width=1.0\textwidth]{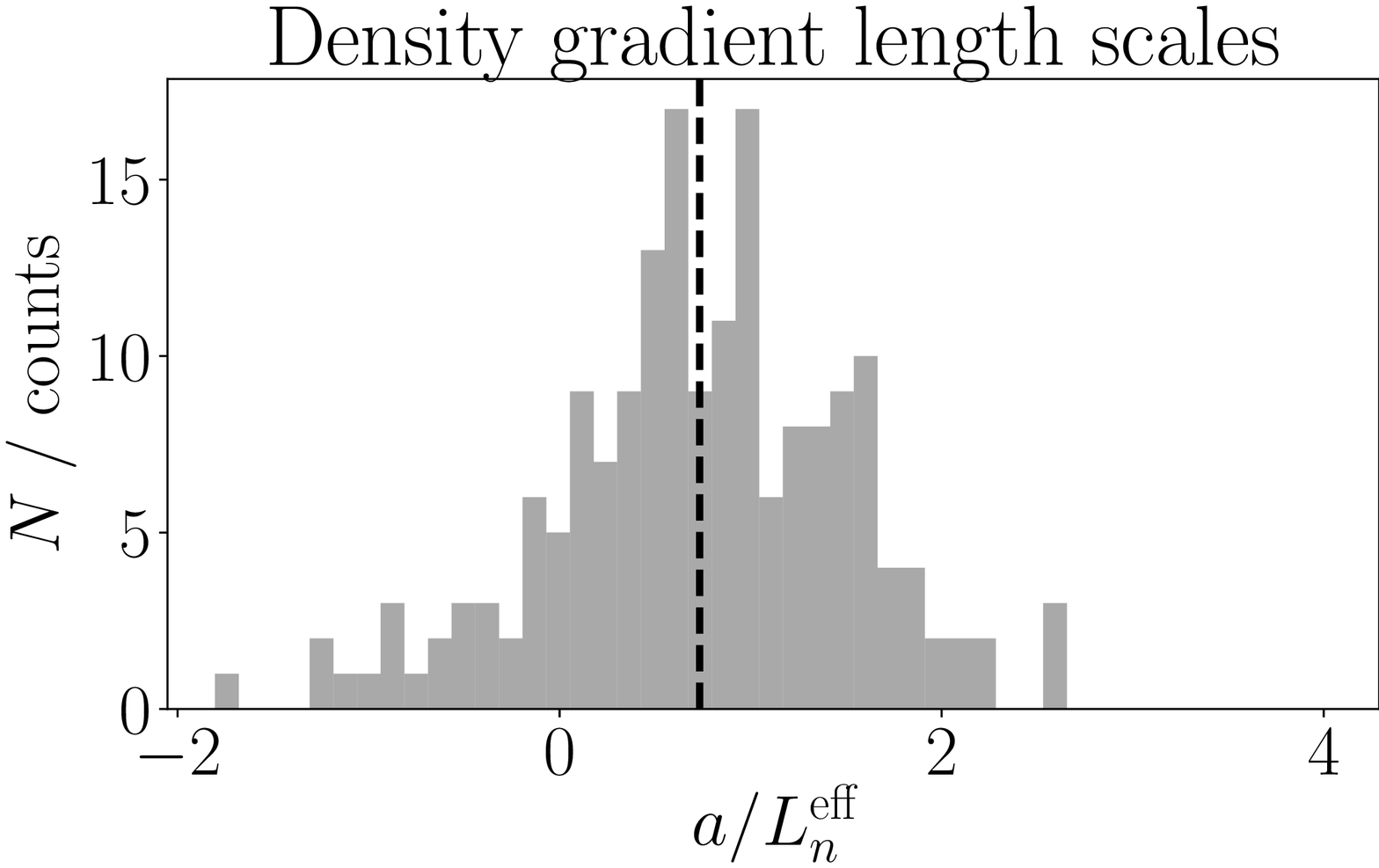}
\caption{} \label{fig:etg_dens_hist_a}
\end{center}
\end{subfigure}

\caption{ (a) The maximum ETG growth rate for CBC magnetic geometry $\growthmaxcbc$
as a function of the normalised background gradient length scales $\lscal/\lte$ and $\lscal/\lln$.
 The cross indicates the nominal CBC values of 
 $\lscal/\lte = 2.3$ and $\lscal/\lln = 0.733$.
(b) The maximum ETG growth rate $\growthmax$ including only the effects
 of the cross-scale term $\evee \cdot \nbls \ihhe{}$ compared to the model growth rate
 $\growthmaxcbc(\lscal/\lteeff,\lscal/\llneffh)$. The dashed line indicates the
 required slope for a perfect correlation.
 (c) $\growthmax$ including the effects
 of the cross-scale term $\evee \cdot \nbls \igge{}$ compared to the model growth rate
 $\growthmaxcbc(\lscal/\lteeff,\lscal/\llneff)$.
 (d)  The histogram of density
 gradient length scales $\lscal/\llneff$. 
  The dashed line indicates $\lscal/\llneff = 0.733$.}
\label{fig:etg_fptp_plots}
\end{figure}
 
The physical interpretation of the impact of $\nbls\igge{}$
 in equation \refeq{equation:electronelectronrealgg} 
 is that $\nbls\igge{}$ modifies the background gradients in $\nbl\eqlbe$ which drive the instability.
 As discussed in section \ref {sec:Interpretation}, $\nbls\igge{}$
 is in general a complicated function of $(\lpar,\energy,\pitch)$.
 In this section we examine to what extent we can model
 $\nbls\igge{}$ by simple modifications to 
 the values of $\lscal/\lte$ and $\lscal/\lln$ locally in the
 IS flux tube.  For this we use the results of the
 simulations presented in figures \ref{fig:etg_gdfnxscale},
 \ref{fig:etg_hdfnxscale} and \ref{fig:etg_phidfnxscale}.
  This simplified model should be valid when the plasma is collisional
  and the fluctuations are zonally dominated: in this limit, $\ihhe{}$ can be
  approximated as a perturbed Maxwellian that has no variation within a flux surface.
 We realise this limiting scenario in section \ref{sec:fptpmarginal}.

 We present the maximum ETG growth rate for the CBC magnetic geometry as
 a function of the background temperature and density gradients,
 $\growthmaxcbc(\lscal/\lte,\lscal/\lln) $, in figure \ref{fig:etg_fptp}.
 We define effective background temperature and density gradients,
      \beqn \frac{\lscal}{\lteeff} = \frac{\lscal}{\lte}
 - \flav{\frac{\lscal \kxfac}{\tempe} \drv{\ith}{\radial} }, \label{eq:lteff} \eeqn 
     \beqn \frac{\lscal}{\llneffh} = \frac{\lscal}{\lln}
     -\flav{ \frac{\lscal\kxfac}{\dens} \drv{\inh}{\radial} },
     \label{eq:llneffh} \eeqn
     and 
     \beqn \frac{\lscal}{\llneff} = \frac{\lscal}{\lln}
     -\flav{ \frac{\lscal\kxfac}{\dens} \drv{\ing}{\radial} },
     \label{eq:llneff} \eeqn     
 respectively, with $\flav{\cdot} = \int^{\pi}_{-\pi} \cdot d \lpar / \bvec\cdot\nbl\lpar$
 an average in $\lpar$ between $(-\pi,\pi)$ at fixed $(\radial,\binormal)$,
 $\inh$ and $\ith$ the density and temperature moments of
 $\ihhe{}$, respectively, and  
  $\ing = \inh + \dens\charge \iptl{} /\tempe$.
 The normalised density gradient scale $\lscal/\llneffh$ measures the effective change
 to $\lscal/\lln$ by density gradients in the long-wavelength nonadiabatic
 response of electrons, whereas $\lscal/\llneff$ measures the effective
 change to $\lscal/\lln$ by gradients in the total long-wavelength density fluctuation. 
 
  In figure \ref{fig:etg_dfn_fptp} we show that there is a strong correlation
  between the $\growthmax$ calculated using equation \refeq{equation:electronelectronrealgg}
  including only the cross-scale term due to the
  IS, nonadiabatic electron response $\evee \cdot \nbls \ihhe{}$ and
  the model growth rate $\growthmaxcbc(\lscal/\lteeff,\lscal/\llneffh)$.
  In contrast, in figure \ref {fig:etg_gdfn_fptp} we show that the cross-scale
  interaction due to $\nbls\igge{}$ cannot be modelled by
  $\growthmaxcbc(\lscal/\lteeff,\lscal/\llneff)$,
  Instead, the model $\growthmaxcbc$ almost always overestimates
  the growth rate of the ETG modes.
 We can understand these results 
 by referring to figure \ref{fig:etg_kparcomparison}.
 Cross-scale interaction due to the electron nonadiabatic response $\nbls \ihhe{}$
 did not change the $\kpara$ spectrum of the ETG eigenmodes: 
 $\nbls \ihhe{}$ has the effect of making the mode more or less unstable by changing the drives only.
 In contrast the difference between $\nbls \igge{}$ and $\nbls\ihhe{}$,
 $\nbls (\charge\iptl{}/\tempe)\eqlbe$, had the effect of introducing larger $\kpara$
 components into the mode, which cannot be captured by simple modifications to
 $\lscal/\lln$ and $\lscal/\lte$. The fact that $\growthmaxcbc(\lscal/\lteeff,\lscal/\llneff)$
 typically overestimates the ETG growth rate in figure \ref{fig:etg_gdfn_fptp} indicates
 that the effect of the parallel-to-the-field variation in $\nbls (\charge\iptl{}/\tempe)\eqlbe$
 is stabilising. 

 Finally, we return to the result presented in figure \ref {fig:etg_gdfnxscale}:
  the average effect of $\nbls \igge{}$ was to suppress ETG instability. We can now understand this
 result qualitatively. Firstly, we note that the CBC value for $\lscal/\lln$
 maximises the ETG growth
 rates at fixed $\lscal/\lte$, as shown in figure \ref{fig:etg_fptp}: any modification
 to $\lscal/\lln$ will reduce the ETG growth rate $\growthmax$.  The histogram of density
 gradient length scales $\lscal/\llneff$ 
 shown in figure \ref {fig:etg_dens_hist_a} indicates that the typical modification
 to $\lscal/\lln$ is sufficient to reduce $\growthmax$ by an amount of order unity. 
 Secondly, the parallel-to-the-field variation in the adiabatic response
 $\nbls (\charge\iptl{}/\tempe)\eqlbe$ suppressed the instability further 
 by increasing the $\kpara$ in the ETG mode.

\begin{figure}
\begin {subfigure} {0.495\textwidth}
\begin{center}
\includegraphics[clip, trim=0cm 1.2cm 0cm 0cm, width=1.0\textwidth]{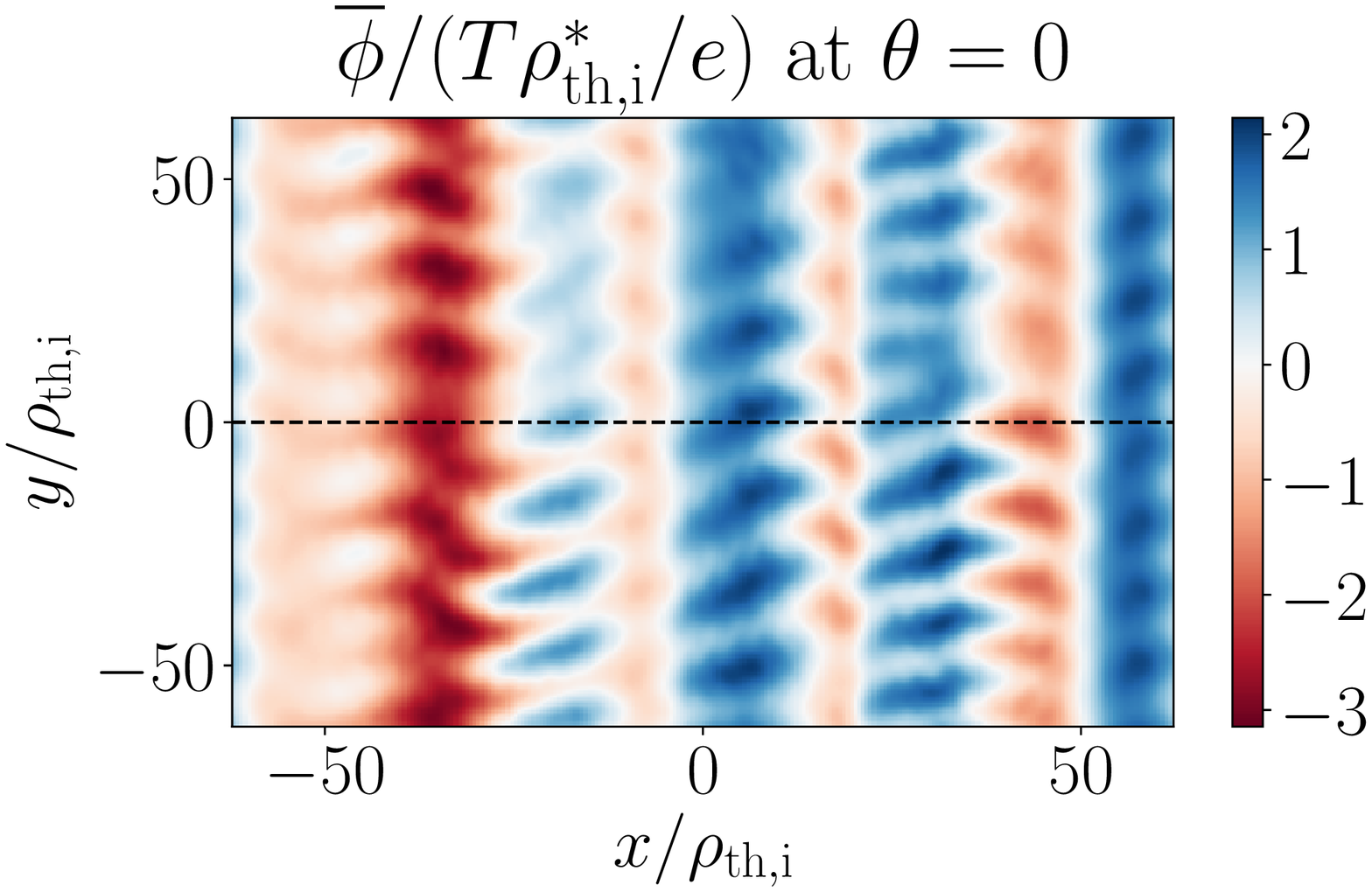}
\caption{} \label{fig:IS_nearmarginal}
\end{center}
\end{subfigure}
\begin{subfigure} {0.495\textwidth}
\begin{center}
\includegraphics[clip, trim=0cm 1.2cm 0cm 0cm, width=1.0\textwidth]{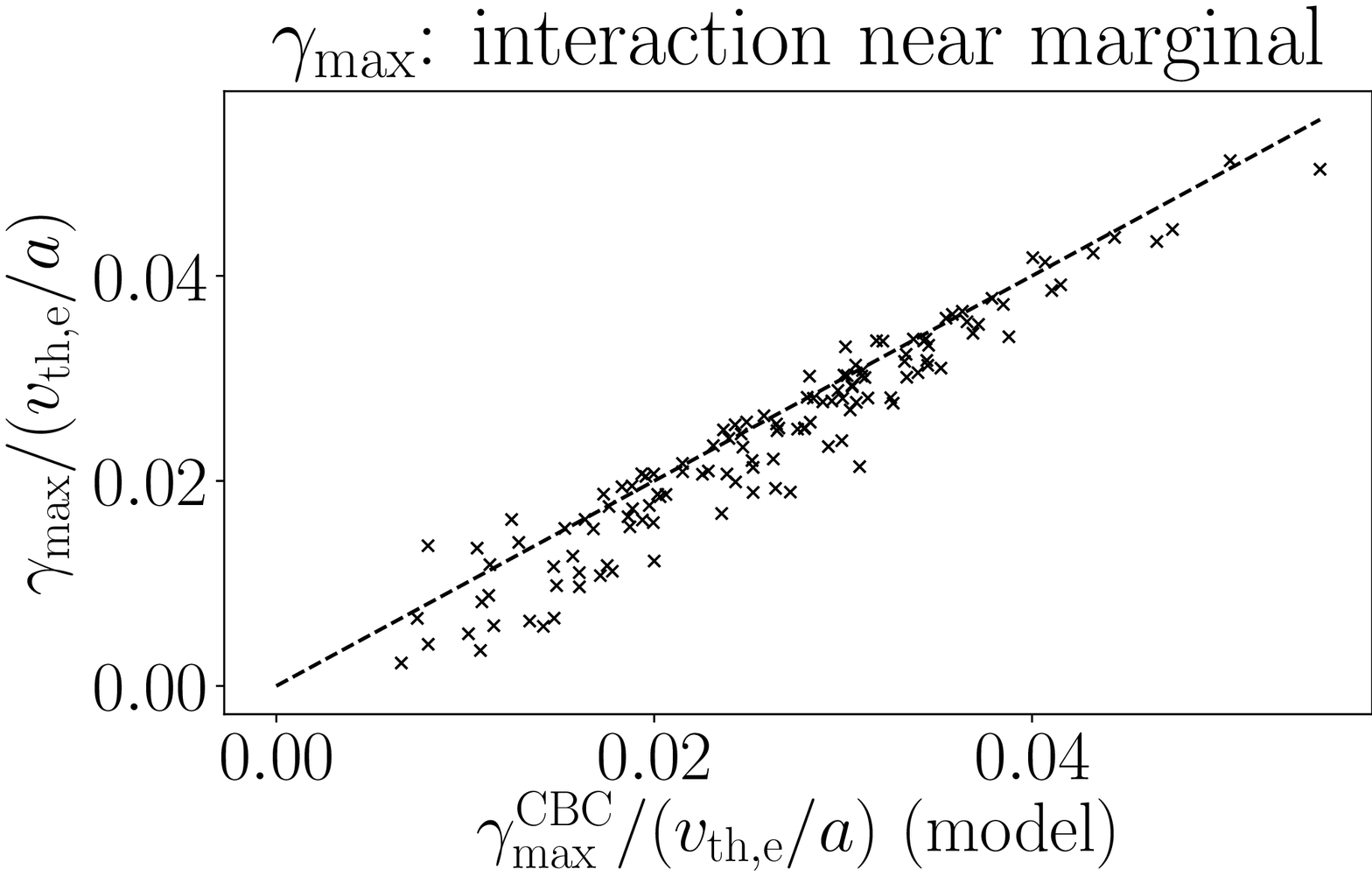}
\caption{} \label{fig:etg_fptp_fullx}
\end{center}
\end{subfigure}

\caption{ (a) Contours of the electrostatic potential $\iptl{}(\radial,\binormal)$
 for the sampled near-marginal IS turbulence.
 The 128 sampled positions $\radials$ lie equally spaced on the dashed line.
(b) The maximum ETG growth rate $\growthmax$ calculated using equation
 \refeq{equation:electronelectronrealgg}, including both cross-scale terms
 $\ivee \cdot \nbls \egge{}$ and $\evee \cdot \nbls \igge{}$, compared to the model growth rate
 $\growthmaxcbc(\lscal/\lteeff,\lscal/\llneff)$
 that is calculated using the results in figure
 \ref{fig:etg_fptp} and definitions \refeq{eq:lteff} and \refeq{eq:llneff}.
 The dashed line indicates the slope for a perfect correlation. }
\label{fig:nearmarginal}
\end{figure}

\section {Cross-scale interaction in near marginal turbulence} \label{sec:fptpmarginal}
The simulation results presented in the previous sections were
 obtained for parameters where there is a strong background
 drive of instability for ITG modes at scales comparable to $\gyrdi$
 and ETG modes at scales comparable to $\gyrde$.
 In this section we present results for 
 the case where $\lscal/\lti$ and $\lscal/\lte$ are reduced so that 
 both the ITG and ETG modes are driven near marginal stability.
 Weakly driven IS turbulence can reach a Dimits-shift
 regime where the 
 fluxes are low, and a long-wavelength,
 long-lived zonal flow regulates the IS dynamics; see, e.g., \cite{DimitsCBCPoP2000,RogersDorland2000}.
 This represents a very different state from strongly driven, critically balanced,
 IS turbulence, cf. \cite{barnesPRL11b}.
 Consequently, the relative importance of parallel-to-the-field shearing and modifications to
 the background drives may differ in turbulence driven near marginal stability.
 
 To obtain a near-marginal case, we use the parameters described in section \ref{sec:Numerical}, 
 but reduce the normalised background temperature gradient length scales
 to $\lscal/\lte = \lscal/\lti =1.38$.
 We use  the resolutions given in appendix \ref{sec:simulationresolutions:nl},
 and we take $\me/\mi =0$ in the IS simulation.
 The IS turbulence is evolved until a long-lived, zonally dominated state emerges.
 We sample the IS gradients $\nbls\igge{}$ and drifts $\ivee$ at 
 radial coordinates in the IS flux tube $\radials$, for a single
 time $\ts$ and binormal position $\binormals$. 
 In figure \ref{fig:IS_nearmarginal} we show $\iptl{}(\radial,\binormal)$ at $\ts$ for
 the near-marginal IS turbulence, with the sampled
 radial points taken along the dashed line.
 At each of the 128 equally spaced $\radials$ positions we perform an ETG linear calculation,
 including the effects of cross-scale interaction via equation \refeq{equation:electronelectronrealgg}
 and using resolutions detailed in appendix \ref{sec:simulationresolutions:lin}.
 In contrast to the strongly driven case, 
 in figure \ref {fig:etg_fptp_fullx} we show that
 the maximum growth rate of the ETG mode
 $\growthmax$ -- calculated including both cross-scale interaction terms, $\ivee \cdot \nbls \egge{}$
 and $\evee \cdot \nbls \igge{}$ --
 is  strongly correlated with 
  the model growth rate $\growthmaxcbc(\lscal/\lteeff,\lscal/\llneff)$, 
 with the effective density gradient scales $\lscal/\lteeff$  and $\lscal/\llneff$
 defined in equations \refeq{eq:lteff} and \refeq{eq:llneff}, respectively,
 and $\growthmaxcbc$ shown in figure \ref{fig:etg_fptp}.
 The 
 imperfections in the correlation in figure
 \ref{fig:etg_fptp_fullx} are primarily due to parallel-to-the-field
 shearing from the term $\ivee \cdot \nblf \egge{}$ in equation \refeq{equation:electronelectronrealgg}
 (cf. figure
 \ref{fig:etg_kparcomparison} which shows that
 $\ivee \cdot \nblf \egge{}$ injects larger $\kpara$ components than
 $\evee \cdot \nblf (\charge\iptl{}/\tempe)\eqlbe$).
 We infer that the effect of parallel-to-the-field
 shearing is weak in this near-marginal turbulence. 
 We verified
 that the average $\kpara$ spectrum of the fastest growing
eigenmodes was identical in the cases with and without cross-scale interaction.
 
 We can understand the result in
 figure \ref{fig:etg_fptp_fullx} with the following intuition.
The wavenumber spectrum of near-marginal turbulence is dominated by the
 zonal mode, a mode which is constant on each flux surface and only depends
 on the radial coordinate. In consequence,
perpendicular gradients in near-marginal turbulence are dominantly in the radial direction,
 and the variation of the the electrostatic potential $\iptl{}$
 in the parallel-to-the-field direction in the turbulence is weak.
 In this regime, 
 the $\exbtext$ advection term $\ivee\cdot\nblf\egge{}$ 
 only modifies the ETG frequency by a Doppler shift \cite{hardmanpaper1}. 
 Hence, the dominant cross-scale interaction term
 that alters the ETG growth rate is due to the radial gradients
 of the IS electron distribution function, i.e.,
 $\nbls\igge{} \simeq \nbl \radial \drvt{\igge{}}{\radial}$.
 In addition, if the plasma is sufficiently collisional then $\igge{}$
 is restricted to be a perturbed Maxwellian of only density and temperature gradients,
 and in consequence, the effect of cross-scale
 interaction in near-marginal turbulence may be parameterised with the effective
 background gradient length scales $\lscal/\lteeff$ and $\lscal/\llneff$.
 The results presented in this section demonstrate that
 this near-marginal cross-scale interaction regime can be realised.

\section{Discussion}\label{sec:discuss} 
 The results presented in this paper were obtained in the limit $\massrt \rightarrow 0 $ 
 with a theory that assumes scale-separation
 between long-wavelength and short-wavelength structures. This theory may not 
 describe multiscale turbulence accurately for
 realistic values of $\massrt$ under a variety of circumstances; e.g., 
 when radially elongated ES structures 
 \cite{dorland2000electron,jenko2000electron, jenko2002prediction} grow to be as large as a
 typical IS eddy; 
 or when there is no significant dissipation at wavelengths between
 the scales of the IS and ES turbulence. 
 Nonetheless, in this paper we have demonstrated that significant cross-scale interactions do 
 persist in the limit $\massrt \rightarrow 0$:
 strongly driven, long-wavelength 
 turbulence can stabilise a strongly driven ETG instability; 
 and long-wavelength turbulence driven near marginal stability
 can change the growth rate of a weakly driven ETG mode by
 an order unity factor.
 The physical mechanisms responsible for these cross-scale interactions were
 shown to be, firstly, the modification of the background drives of instability by gradients
 in long-wavelength fluctuations, and secondly, the parallel-to-the-field shearing of
 short-wavelength fluctuations by long-wavelength $\exbtext$ drifts. 
 In the case of strongly driven turbulence, the effects of 
 both parallel-to-the-field shearing and nonuniform modified background drives
 contributed to the stabilisation of the ETG instability.
However, in the case of near-marginal turbulence, the
 effect of parallel-to-the-field shearing was weak,
 and the effect of cross-scale interaction could be
 parameterised with the effective background
 gradient length scales $\lscal/\llneff$ and $\lscal/\lteeff$.
 
 We can qualitatively explain the difference in the nature of the cross-scale interactions
 that we find in the cases of near-marginal and strongly driven turbulence.
 Near-marginal ITG turbulence is dominated by zonal modes, see e.g., \cite{DimitsCBCPoP2000,RogersDorland2000}.
 Zonal modes have a structure that depends only on radial position, and hence 
 the long-wavelength $\exbtext$ drift $\ivee$ only has the effect of 
 shifting the ETG fluctuation frequency by a Doppler shift -- for zonal modes where $\iptl{} = \iptl{}(\radial)$
 the term $\ivee \cdot \nblf \egge{}$ in equation \refeq{equation:electronelectronrealgg}
 can be removed by boosting to a toroidally rotating frame \cite{mhardman2019a}.
 In this situation where the effect of
 parallel shearing is negligible, the dominant cross-scale interaction
 arises from the radial gradient $\nbls\igge{}\simeq \nbl \radial \drvt{\igge{}}{\radial}$.
 If $\igge{}$ is sufficiently close to a perturbed Maxwellian
 -- perhaps as a result of moderate collisionality
 -- then we find that $\drvt{\igge{}}{\radial}$ has dominant contributions from a density gradient and a temperature gradient. 
 In contrast, strongly driven ITG turbulence is not zonally dominated,
 but consists of critically balanced turbulent eddies \cite {barnesPRL11b}
 that can have a parallel correlation length as large as the device scale $\saffac \rmaj$.
 Hence, $\nbls\igge{}$  and $\ivee$ can contain parallel-to-the-field variation
 that results in the nonuniform drive and the shearing apart of short-wavelength modes, respectively. 
 In critically balanced turbulence, eddies with a
 greater $\ekky$ have a faster nonlinear
 turnover time, and hence a shorter parallel correlation length 
 and a greater characteristic $\kpara\propto \ekky^{4/3}(\rmaj/\lti)^{4/3}$ \cite{barnesPRL11b}.
 A range of $\ekky$ modes in the inertial range can contribute to the
 gradient $\nbls\igge{}$ and the drift $\ivee$, 
 with the result that $\nbls\igge{}$ and $\ivee$ can have
 parallel-to-the-field oscillations on scales shorter than $\saffac\rmaj$.
 This observation leads us to expect that the
 impact of parallel-to-the-field shearing increases
 with increasing $\rmaj/\lti$.
 
 A criterion for when to expect parallel-to-the-field $\exbtext$ shear
 to suppress ETG modes can be obtained from a simple
 quasilinear argument. 
 The effective shearing rate $\whate$ should scale 
 with the parallel-to-the-field variation in the
 cross-scale advection operator $\ivee\cdot\nblf$, i.e.,
 $\whate \sim \ekkest \drvt {\ivee}{ \lpar}$.
 We can expect
 parallel-to-the-field shear suppression when the effective shear rate
 \beqn \whate  \gtrsim \egammaest; \label{eq:effw1}\eeqn i.e., 
 when the shear rate is faster than the typical ETG mode growth rate $\egammaest$,
 at a typical ETG mode wave number $\ekkest$.
 A quasilinear estimate gives
 $\ivee \sim \igammaest / \ikkest$, with $\igammaest$ and $\ikkest$ the typical ITG mode growth rate
 and wave number, respectively. 
  If we assume that the variation of $\ivee$
 in the parallel-to-the-field direction $\drvt{ \ivee }{\lpar} \sim \ivee$, then
 we find that we can expect parallel-to-the-field
 shear suppression of ETG turbulence when 
 \beqn \frac{\igammaest}{\ikkest} \gtrsim  \frac{\egammaest}{\ekkest}.\label{eq:effw2} \eeqn 
 This result is consistent with observations of the behaviour
 of multiscale turbulence made in light of some DNS
 \cite{2016StaeblerMultiscale,2017StaeblerMultiscale,Creely_2019},
 though their interpretation differs from the one given here:
 \cite{2016StaeblerMultiscale,2017StaeblerMultiscale,Creely_2019}
 neglect the variation of $\exbtext$
 drifts along field lines, and so neglect the key physical mechanism
  that is critical to the quasilinear argument:
 the effect of parallel-to-the-field $\exbtext$ shear. 
 We note that parallel-to-the-field flow shear has previously been considered as a part
  of a model of hyperviscous dissipation of ITG driven turbulence \cite{SmithThesis}.
 
 The suppression of short-wavelength turbulence seen in some DNS is often assumed to be the result 
 of perpendicular-to-the-field $\exbtext$ shearing  by
 long-wavelength turbulence.
 In fact, that mechanism
 does not appear in the leading-order, scale-separated ES equation
 \refeq{equation:electronelectronrealgg} \cite{hardmanpaper1}. 
 In the limit that $\massrt \rightarrow 0$,
 ES structures do not have a large enough spatial extent 
 to be affected by perpendicular-to-the-field shear
 in IS $\exbtext$ flows. 
 
 Finally, we note that the scale-separated model should be modified
 if factors that we have assumed to be of order unity become
 large enough to interfere with the $\massrt$ expansion;
 possible examples of such parameters include
 the ratio $(\egammaest/\ekkest)/(\igammaest/\ikkest)$,
 the degree of spatial anisotropy in the turbulence,
 and the ratio of the zonal to nonzonal fluctuation amplitudes
  \cite{hardmanpaper1}.

 \par \textit{
 The authors would like to thank A. A. Schekochihin, W. Dorland, F. I. Parra, P. Dellar, S. C. Cowley,
 J. Ball, A. Geraldini, N. Christen, A. Mauriya, P. Ivanov, Y. Kawazura, P. Ivanov, J. Parisi, 
  and J. Ruiz Ruiz 
 for useful discussion.
 This work has been carried out within the framework of the EUROfusion Consortium and
 has received funding from the Euratom research and training programme 2014-2018 and 2019-2020 under grant
 agreement No 633053 and from the RCUK Energy Programme [Grant Numbers EP/P012450/1 and EP/T012250/1].
 The views and opinions expressed herein do not necessarily reflect those of the European Commission.
 This work was supported in part by the Engineering and Physical Sciences Research Council (EPSRC) [Grant Number EP/R034737/1].
 The author acknowledges EUROfusion, the EUROfusion High Performance Computer (Marconi-Fusion) under the project MULTEI,
 the use of ARCHER under the project e607, and the use of ARCHER through the
 Plasma HEC Consortium EPSRC Grant Numbers EP/L000237/1 and EP/R029148/1 under the projects e281-gs2, 
 software support 
 through the Plasma-CCP Network under EPSRC Grant Number 
  EP/M022463/1, and support from the Wolfgang Pauli Institute.}

\appendix

\section {Simulation Resolutions}\label{sec:simulationresolutions}
    This appendix details the resolutions used for the simulations presented in this paper.
    \subsection{Nonlinear simulations}\label{sec:simulationresolutions:nl}
    
    For the nonlinear simulations presented in section \ref{sec:Numerical} we use
    a flux tube with a perpendicular cross-section of $40\pi \gyrdi \times 40\pi \gyrdi$,
    and a single poloidal turn in the parallel-to-the-field direction.
    We use $\noy = 21$ toroidal modes with $\kky > 0$ and  $\ikkymax \gyrdi = 1.05$.
    We use $\nox = 255$ radial modes with   $\ikkxmax \gyrdi = 6.38$.
    We specify that each $(\kkx,\kky)$ mode has a $[-\pi,\pi]$ extent
    in $\lpar$, with $\nlpar =33$ points.
    We use a $\pitch$ grid with $\npitch = 27$.
    To describe passing particles we take $\npitchp = 11$ points,
    chosen using Gauss-Radau rules \cite{Hildebrandnumericalanalysis}.    
    For the trapped particles we employ a nonspectral,
    unequally spaced grid in $\pitch$, with $\npitcht = 16$ points
    at the outboard midplane.
    Following \cite{barnesPoP10a}, we use an $\energy$ grid derived from a 
    spectral speed $\vmag = \sqrt{2 \energy/ \mass}$ grid with $\negrid = 12$ points.
    For the near-marginal nonlinear simulation presented in section \ref{sec:fptpmarginal}
    we use the resolutions described above, with $\negrid = 16$.
    
    \subsection{Linear simulations}\label{sec:simulationresolutions:lin}
    
    For the linear simulations presented in sections \ref{sec:Numerical} - \ref{sec:fptp}
    we use ballooning modes with an extent $[-3\pi,3\pi]$ in the ballooning angle.
    For compatibility with the nonlinear simulations, each $[-\pi,\pi]$ segment has $\nlpar=33$  points,
    and we take $\npitch = 27$ and $\negrid =12$.    
    For the near-marginal linear simulations presented in section \ref{sec:fptpmarginal}
    we use the linear resolutions described above, with $\negrid = 16$ for
    consistency with the near-marginal nonlinear simulation.

\bibliography{stabilisation_of_short_wavelength_instabilities_main}

\begin{thebibliography}{37}
\expandafter\ifx\csname natexlab\endcsname\relax\def\natexlab#1{#1}\fi
\def\au#1{#1} \def\ed#1{#1} \def\yr#1{#1}\def\at#1{#1}\def\jt#1{\textit{#1}}
  \def\bt#1{#1}\def\bvol#1{\textbf{#1}} \def\vol#1{#1} \def\pg#1{#1}
  \def\publ#1{#1}\def\arxiv#1{#1}\def\org#1{#1}\def\st#1{\textit{#1}}

\bibitem[Abel {\em et~al.\/}(2013)Abel, Plunk, Wang, Barnes, Cowley, Dorland \&
  Schekochihin]{abelRPP13}
{\sc \au{Abel, I.~G.}, \au{Plunk, G.~G.}, \au{Wang, E.}, \au{Barnes, M.},
  \au{Cowley, S.~C.}, \au{Dorland, W.} \& \au{Schekochihin, A.~A.}} \yr{2013}
  \at{Multiscale gyrokinetics for rotating tokamak plasmas: fluctuations,
  transport, and energy flows}.  \jt{Reports on Progress in Physics}
  \bvol{76},  \pg{116201}.

\bibitem[Barnes {\em et~al.\/}(2010)Barnes, Dorland \& Tatsuno]{barnesPoP10a}
{\sc \au{Barnes, M.}, \au{Dorland, W.} \& \au{Tatsuno, T.}} \yr{2010}
  \at{Velocity space resolution in gyrokinetic simulations}.  \jt{Phys.
  Plasmas}  \bvol{17},  \pg{032106}.

\bibitem[Barnes {\em et~al.\/}(2011)Barnes, Parra \&
  Schekochihin]{barnesPRL11b}
{\sc \au{Barnes, M.}, \au{Parra, F.~I.} \& \au{Schekochihin, A.~A.}} \yr{2011}
  \at{Critically balanced ion temperature gradient turbulence in fusion
  plasmas}.  \jt{Phys. Rev. Lett.}  \bvol{107},  \pg{115003}, arxiv:1104.4514.

\bibitem[Bonanomi {\em et~al.\/}(2018)Bonanomi, Mantica, Citrin, Goerler \&
  and]{Bonanomi_2018_ImpactofES}
{\sc \au{Bonanomi, N.}, \au{Mantica, P.}, \au{Citrin, J.}, \au{Goerler, T.} \&
  \au{and, B.~T.}} \yr{2018}  \at{Impact of electron-scale turbulence and
  multi-scale interactions in the {JET} tokamak}.  \jt{Nuclear Fusion}
  \bvol{58},  \pg{124003}.

\bibitem[Brizard \& Hahm(2007)]{brizardRMP07}
{\sc \au{Brizard, A.~J.} \& \au{Hahm, T.~S.}} \yr{2007}  \at{Foundations of
  nonlinear gyrokinetic theory}.  \jt{Rev. Mod. Phys.}  \bvol{79},  \pg{421}.

\bibitem[Candy {\em et~al.\/}(2007)Candy, Waltz, Fahey \&
  Holland]{candy2007effect}
{\sc \au{Candy, J.}, \au{Waltz, R.~E.}, \au{Fahey, M.~R.} \& \au{Holland, C.}}
  \yr{2007}  \at{The effect of ion-scale dynamics on
  electron-temperature-gradient turbulence}.  \jt{Plasma Physics and Controlled
  Fusion}  \bvol{49},  \pg{1209}.

\bibitem[Catto(1978)]{cattoPP78}
{\sc \au{Catto, P.~J.}} \yr{1978}  \at{Linearized gyro-kinetics}.  \jt{Plasma
  Phys.}  \bvol{20},  \pg{719}.

\bibitem[Cowley {\em et~al.\/}(1991)Cowley, Kulsrud \& Sudan]{cowleyPoFB91}
{\sc \au{Cowley, S.~C.}, \au{Kulsrud, R.~M.} \& \au{Sudan, R.}} \yr{1991}
  \at{Considerations of ion-temperature-gradient-driven turbulence}.  \jt{Phys.
  Fluids B}  \bvol{3},  \pg{2767}.

\bibitem[Creely {\em et~al.\/}(2019)Creely, Rodriguez-Fernandez, Conway,
  Freethy, Howard \& and]{Creely_2019}
{\sc \au{Creely, A.~J.}, \au{Rodriguez-Fernandez, P.}, \au{Conway, G.~D.},
  \au{Freethy, S.~J.}, \au{Howard, N.~T.} \& \au{and, A. E.~W.}} \yr{2019}
  \at{Criteria for the importance of multi-scale interactions in turbulent
  transport simulations}.  \jt{Plasma Physics and Controlled Fusion}
  \bvol{61},  \pg{085022}.

\bibitem[Dimits {\em et~al.\/}(2000)Dimits, Bateman, Beer, Cohen, Dorland,
  Hammett, Kim, Kinsey, Kotschenreuther, Kritz, Lao, Mandrekas, Nevins, Parker,
  Redd, Shumaker, Sydora \& Weiland]{DimitsCBCPoP2000}
{\sc \au{Dimits, A.~M.}, \au{Bateman, G.}, \au{Beer, M.~A.}, \au{Cohen, B.~I.},
  \au{Dorland, W.}, \au{Hammett, G.~W.}, \au{Kim, C.}, \au{Kinsey, J.~E.},
  \au{Kotschenreuther, M.}, \au{Kritz, A.~H.}, \au{Lao, L.~L.}, \au{Mandrekas,
  J.}, \au{Nevins, W.~M.}, \au{Parker, S.~E.}, \au{Redd, A.~J.}, \au{Shumaker,
  D.~E.}, \au{Sydora, R.} \& \au{Weiland, J.}} \yr{2000}  \at{Comparisons and
  physics basis of tokamak transport models and turbulence simulations}.
  \jt{Physics of Plasmas}  \bvol{7},  \pg{969--983}.

\bibitem[Dorland {\em et~al.\/}(2000)Dorland, Jenko, Kotschenreuther \&
  Rogers]{dorland2000electron}
{\sc \au{Dorland, W.}, \au{Jenko, F.}, \au{Kotschenreuther, M.} \& \au{Rogers,
  B.~N.}} \yr{2000}  \at{Electron temperature gradient turbulence}.  \jt{Phys.
  Rev. Lett.}  \bvol{85},  \pg{5579--5582}.

\bibitem[Frieman \& Chen(1982)]{friemanPoF82}
{\sc \au{Frieman, E.~A.} \& \au{Chen, L.}} \yr{1982}  \at{Nonlinear gyrokinetic
  equations for low-frequency electromagnetic waves in general plasma
  equilibria}.  \jt{Phys. Fluids}  \bvol{25},  \pg{502}.

\bibitem[G\"{o}rler \& Jenko(2008)]{gorler2008scale}
{\sc \au{G\"{o}rler, T.} \& \au{Jenko, F.}} \yr{2008}  \at{Scale separation
  between electron and ion thermal transport}.  \jt{Physical review letters}
  \bvol{100},  \pg{185002}.

\bibitem[Hardman(2019)]{mhardman2019a}
{\sc \au{Hardman, M.}} \yr{2019}  \at{Multiscale turbulence in magnetic
  confinement fusion devices, {PhD} {Thesis}}.  \jt{University of Oxford} .

\bibitem[Hardman {\em et~al.\/}(2019)Hardman, Barnes, Roach \&
  Parra]{hardmanpaper1}
{\sc \au{Hardman, M.~R.}, \au{Barnes, M.}, \au{Roach, C.~M.} \& \au{Parra,
  F.~I.}} \yr{2019}  \at{A scale-separated approach for studying coupled ion
  and electron scale turbulence}.  \jt{Plasma Physics and Controlled Fusion}
  \bvol{61},  \pg{065025}.

\bibitem[Hildebrand(1987)]{Hildebrandnumericalanalysis}
{\sc \au{Hildebrand, F.~B.}} \yr{1987} {\em Introduction to Numerical Analysis,
  Second edition\/}.  \publ{Dover, New York}.

\bibitem[Horton {\em et~al.\/}(1988)Horton, Hong \& Tang]{HortonETG}
{\sc \au{Horton, W.}, \au{Hong, B.~G.} \& \au{Tang, W.~M.}} \yr{1988}
  \at{Toroidal electron temperature gradient driven drift modes}.  \jt{The
  Physics of Fluids}  \bvol{31},  \pg{2971--2983}.

\bibitem[Howard {\em et~al.\/}(2014)Howard, Holland, White, Greenwald \&
  Candy]{howard2014synergistic}
{\sc \au{Howard, N.~T.}, \au{Holland, C.}, \au{White, A.~E.}, \au{Greenwald,
  M.} \& \au{Candy, J.}} \yr{2014}  \at{Synergistic cross-scale coupling of
  turbulence in a tokamak plasma}.  \jt{Physics of Plasmas}  \bvol{21}.

\bibitem[Howard {\em et~al.\/}(2015)Howard, Holland, White, Greenwald \&
  Candy]{howard2015fidelity}
{\sc \au{Howard, N.~T.}, \au{Holland, C.}, \au{White, A.~E.}, \au{Greenwald,
  M.} \& \au{Candy, J.}} \yr{2015}  \at{Fidelity of reduced and realistic
  electron mass ratio multi-scale gyrokinetic simulations of tokamak
  discharges}.  \jt{Plasma Physics and Controlled Fusion}  \bvol{57},
  \pg{065009}.

\bibitem[Howard {\em et~al.\/}(2016{\natexlab{{\em a\/}}})Howard, Holland,
  White, Greenwald \& Candy]{howard2016enhanced}
{\sc \au{Howard, N.~T.}, \au{Holland, C.}, \au{White, A.~E.}, \au{Greenwald,
  M.} \& \au{Candy, J.}} \yr{2016{\natexlab{{\em a\/}}}}  \at{Multi-scale
  gyrokinetic simulation of tokamak plasmas: {E}nhanced heat loss due to
  cross-scale coupling of plasma turbulence}.  \jt{Nuclear Fusion}  \bvol{56},
  \pg{014004}.

\bibitem[Howard {\em et~al.\/}(2016{\natexlab{{\em b\/}}})Howard, Holland,
  White, Greenwald, Candy \& Creely]{howard2016comparison}
{\sc \au{Howard, N.~T.}, \au{Holland, C.}, \au{White, A.~E.}, \au{Greenwald,
  M.}, \au{Candy, J.} \& \au{Creely, A.~J.}} \yr{2016{\natexlab{{\em b\/}}}}
  \at{Multi-scale gyrokinetic simulations: {C}omparison with experiment and
  implications for predicting turbulence and transport}.  \jt{Physics of
  Plasmas}  \bvol{23},  \pg{056109}.

\bibitem[Itoh \& Itoh(2001)]{Itoh_2001}
{\sc \au{Itoh, S.-I.} \& \au{Itoh, K.}} \yr{2001}  \at{Statistical theory and
  transition in multiple-scale-length turbulence in plasmas}.  \jt{Plasma
  Physics and Controlled Fusion}  \bvol{43},  \pg{1055--1102}.

\bibitem[Jenko \& Dorland(2002)]{jenko2002prediction}
{\sc \au{Jenko, F.} \& \au{Dorland, W.}} \yr{2002}  \at{Prediction of
  significant tokamak turbulence at electron gyroradius scales}.  \jt{Phys.
  Rev. Lett.}  \bvol{89},  \pg{225001}.

\bibitem[Jenko {\em et~al.\/}(2000)Jenko, Dorland, Kotschenreuther \&
  Rogers]{jenko2000electron}
{\sc \au{Jenko, F.}, \au{Dorland, W.}, \au{Kotschenreuther, M.} \& \au{Rogers,
  B.~N.}} \yr{2000}  \at{Electron temperature gradient driven turbulence}.
  \jt{Physics of Plasmas}  \bvol{7},  \pg{1904--1910}.

\bibitem[Kotschenreuther {\em et~al.\/}(1995)Kotschenreuther, Rewoldt \&
  Tang]{KOTSCHENREUTHER1995CPC}
{\sc \au{Kotschenreuther, M.}, \au{Rewoldt, G.} \& \au{Tang, W.}} \yr{1995}
  \at{Comparison of initial value and eigenvalue codes for kinetic toroidal
  plasma instabilities}.  \jt{Computer Physics Communications}  \bvol{88},
  \pg{128 -- 140}.

\bibitem[Lee {\em et~al.\/}(1987)Lee, Dong, Guzdar \& Liu]{LeeETG}
{\sc \au{Lee, Y.~C.}, \au{Dong, J.~Q.}, \au{Guzdar, P.~N.} \& \au{Liu, C.~S.}}
  \yr{1987}  \at{Collisionless electron temperature gradient instability}.
  \jt{The Physics of Fluids}  \bvol{30},  \pg{1331--1339}.

\bibitem[Maeyama {\em et~al.\/}(2015)Maeyama, Idomura, Watanabe, Nakata, Yagi,
  Miyato, Ishizawa \& Nunami]{maeyama2015cross}
{\sc \au{Maeyama, S.}, \au{Idomura, Y.}, \au{Watanabe, T.-H.}, \au{Nakata, M.},
  \au{Yagi, M.}, \au{Miyato, N.}, \au{Ishizawa, A.} \& \au{Nunami, M.}}
  \yr{2015}  \at{Cross-scale interactions between electron and ion scale
  turbulence in a tokamak plasma}.  \jt{Physical review letters}  \bvol{114},
  \pg{255002}.

\bibitem[Maeyama {\em et~al.\/}(2017{\natexlab{{\em a\/}}})Maeyama, Watanabe,
  Idomura, Nakata, Ishizawa \& Nunami]{Maeyama_2017_NF}
{\sc \au{Maeyama, S.}, \au{Watanabe, T.-H.}, \au{Idomura, Y.}, \au{Nakata, M.},
  \au{Ishizawa, A.} \& \au{Nunami, M.}} \yr{2017{\natexlab{{\em a\/}}}}
  \at{Cross-scale interactions between turbulence driven by electron and ion
  temperature gradients via sub-ion-scale structures}.  \jt{Nuclear Fusion}
  \bvol{57},  \pg{066036}.

\bibitem[Maeyama {\em et~al.\/}(2017{\natexlab{{\em b\/}}})Maeyama, Watanabe \&
  Ishizawa]{maeyama2017supression}
{\sc \au{Maeyama, S.}, \au{Watanabe, T.-H.} \& \au{Ishizawa, A.}}
  \yr{2017{\natexlab{{\em b\/}}}}  \at{Suppression of ion-scale microtearing
  modes by electron-scale turbulence via cross-scale nonlinear interactions in
  tokamak plasmas}.  \jt{Phys. Rev. Lett.}  \bvol{119},  \pg{195002}.

\bibitem[Parisi {\em et~al.\/}(2020)Parisi, Parra, Roach, Giroud, Dorland,
  Hatch, Barnes, Hillesheim, Aiba, J.~Ball \& Contributors]{Parisi_arXiv_2020a}
{\sc \au{Parisi, J.~F.}, \au{Parra, F.~I.}, \au{Roach, C.~M.}, \au{Giroud, C.},
  \au{Dorland, W.}, \au{Hatch, D.~R.}, \au{Barnes, M.}, \au{Hillesheim, J.~C.},
  \au{Aiba, N.}, \au{J.~Ball, P. G.~I.} \& \au{Contributors, J.}} \yr{2020}
  \at{Toroidal and slab {ETG} instability dominance in the linear spectrum of
  {JET-ILW} pedestals}.  \jt{arXiv} ArXiv:2004.13634.

\bibitem[Rogers {\em et~al.\/}(2000)Rogers, Dorland \&
  Kotschenreuther]{RogersDorland2000}
{\sc \au{Rogers, B.~N.}, \au{Dorland, W.} \& \au{Kotschenreuther, M.}}
  \yr{2000}  \at{Generation and stability of zonal flows in
  ion-temperature-gradient mode turbulence}.  \jt{Phys. Rev. Lett.}  \bvol{85},
   \pg{085022}.

\bibitem[Romanelli(1989)]{RomanelliITG}
{\sc \au{Romanelli, F.}} \yr{1989}  \at{Ion temperature-gradient-driven modes
  and anomalous ion transport in tokamaks}.  \jt{Physics of Fluids B: Plasma
  Physics}  \bvol{1},  \pg{1018--1025}.

\bibitem[Smith(1997)]{SmithThesis}
{\sc \au{Smith, S.~A.}} \yr{1997}  \at{Dissipative closures for statistical
  moments, fluid moments, and subgrid scales in plasma turbulence.}  \jt{PhD
  thesis, Princeton University} .

\bibitem[Staebler {\em et~al.\/}(2016)Staebler, Candy, Howard \&
  Holland]{2016StaeblerMultiscale}
{\sc \au{Staebler, G.~M.}, \au{Candy, J.}, \au{Howard, N.~T.} \& \au{Holland,
  C.}} \yr{2016}  \at{The role of zonal flows in the saturation of multi-scale
  gyrokinetic turbulence}.  \jt{Physics of Plasmas}  \bvol{23},  \pg{062518}.

\bibitem[Staebler {\em et~al.\/}(2017)Staebler, Howard, Candy \&
  Holland]{2017StaeblerMultiscale}
{\sc \au{Staebler, G.~M.}, \au{Howard, N.~T.}, \au{Candy, J.} \& \au{Holland,
  C.}} \yr{2017}  \at{A model of the saturation of coupled electron and ion
  scale gyrokinetic turbulence}.  \jt{Nuclear Fusion}  \bvol{57},  \pg{066046}.

\bibitem[Sugama \& Horton(1997)]{sugamaPoP97}
{\sc \au{Sugama, H.} \& \au{Horton, W.}} \yr{1997}  \at{Transport processes and
  entropy production in toroidally rotating plasmas with electrostatic
  turbulence}.  \jt{Phys. Plasmas}  \bvol{4},  \pg{405}.

\bibitem[Waltz {\em et~al.\/}(2007)Waltz, Candy \& Fahey]{waltz2007coupled}
{\sc \au{Waltz, R.~E.}, \au{Candy, J.} \& \au{Fahey, M.}} \yr{2007}
  \at{Coupled ion temperature gradient and trapped electron mode to electron
  temperature gradient mode gyrokinetic simulations a}.  \jt{Physics of
  plasmas}  \bvol{14},  \pg{056116}.

\end{thebibliography}

\end{document}
%